\documentclass[pra,twocolumn,showpacs,eqsecnum,superscriptaddress]{revtex4}

\usepackage{graphicx}
\usepackage[urlcolor=blue, hyperindex, colorlinks, bookmarks=true,linkcolor=black,citecolor=black]{hyperref}

\begin{document}
\vfuzz2pt 
\hfuzz2pt 

\newcommand{\sch}{Schr\"odinger~}
\newcommand{\schs}{Schr\"odinger's~}
\newcommand{\nn}{\nonumber}
\newcommand{\nl}{\nn \\ &&}
\newcommand{\dg}{^\dagger}
\newcommand{\rt}[1]{\sqrt{#1}\,}
\newcommand{\bra}[1]{\langle{#1}|}
\newcommand{\ket}[1]{|{#1}\rangle}
\newcommand{\ito}{It\^o~}
\newcommand{\str}{Stratonovich~}
\newcommand{\erf}[1]{Eq.~(\ref{#1})}
\newcommand{\erfs}[2]{Eqs.~(\ref{#1}) and (\ref{#2})}
\newcommand{\erft}[2]{Eqs.~(\ref{#1}) -- (\ref{#2})}

\newcommand{\bfr}{{\bf R}}
\newcommand{\bff}{{\bf F}}
\newcommand{\bfrf}{{\bf R, F}}
\newcommand{\smallfrac}[2]{\mbox{$\frac{#1}{#2}$}}
\newcommand{\rhoc}{\rho_{{ r}}}
\newcommand{\ip}[2]{\left\langle{#1}\right|\left.{#2}\right\rangle}
\newcommand{\sq}[1]{\left[ {#1} \right]}
\newcommand{\rhoi}{\rho_{{\bf R}}}
\newcommand{\rhoif}{\rho_{{\bf R, F}}}
\newcommand{\upsik}{\ket{\tilde{\psi}_{{\bf R, F}}(t)}}
\newcommand{\upsib}{\bra{\tilde{\psi}_{{\bf R, F}}(t)}}
\newcommand{\psik}{\ket{{\psi}_{{\bf R, F}}(t)}}
\newcommand{\psib}{\bra{{\psi}_{{\bf R, F}}(t)}}
\newcommand{\lpsik}{\ket{\bar{\psi}_{{\bf R, F}}(t)}}
\newcommand{\lpsib}{\bra{\bar{\psi}_{{\bf R, F}}(t)}}

\title{Stochastic simulations of conditional states of partially observed systems, quantum and classical}

\date{\today}
\author{Jay Gambetta} 
\affiliation{Department of
Physics, Yale University, New Haven, CT 06520}
\affiliation{Centre for Computer Technology, Centre
for Quantum Dynamics, School of Science, Griffith University,
Brisbane 4111, Australia}
\author{H. M. Wiseman} \email{h.wiseman@griffith.edu.au} 
\affiliation{Centre for Computer Technology, Centre
for Quantum Dynamics, School of Science, Griffith University,
Brisbane 4111, Australia}


\begin{abstract}
In a partially observed quantum or classical system the
information that we cannot access results in our description of
the system becoming mixed, even if we have perfect initial
knowledge. That is, if the system is quantum the conditional state
will be given by a state matrix $\rho_r(t)$, and if classical, the
conditional state will be given by a probability distribution
$P_r(x,t)$, where $r$ is the result of the measurement. Thus to
determine the evolution of this conditional state, under
continuous-in-time monitoring, requires a numerically expensive
calculation. In this paper we demonstrate a numerical
technique based on linear measurement theory that allows us to
determine the conditional state using only pure states. That is,
our technique reduces the problem size by a factor of $N$, the number
of basis states for the system. Furthermore we show that our
method can be applied to joint classical and quantum systems such as 
 arise in modeling realistic (finite bandwidth, noisy) measurement.
\end{abstract}


\maketitle

\section{Introduction}

To obtain information about a system, a measurement has to be made.
Based on the results of this measurement we assign to the
system our state of knowledge. For a classical system this state
takes the form of a probability distribution $P(x',t)$, while for a
quantum system we have a state matrix $\rho(t)$. \footnote{Here we are not
concerned with where the division between classical systems and
quantum systems occurs.  Instead we recognize that both
descriptions are valid and the system dynamics determine which is
appropriate.} In this paper we are concerned with efficient
simulation techniques for {\em partly} observed systems; that is,
systems for which the observer cannot obtain enough information to
assign the system a pure state, $P(x',t)=\delta[x'-x(t)]$ or
$\rho(t)=\ket{\psi(t)}\bra{\psi(t)}$.

The chief motivation for wishing to know the conditional state of a system
is for the purpose of feedback control \cite{Jac93,Bel87,WisMil94,DohJac99}. That is because
for cost functions that are additive in time, the optimal basis for controlling the
system is the observer's state of knowledge about the system. Even if such a control strategy
is too difficult to implement in practice, it plays the important role of bounding the
performance of any strategy, which helps in seeking the best practical strategy.

It is well known that the quantum state of an open quantum system, 
given continuous-in-time measurements of the bath, follows a
stochastic trajectory through time \cite{BelSta92}. In the quantum
optics community this is referred to as a quantum trajectory
\cite{Car93a,GarParZol92,MolCasDal93,WisMil93a,GoeGra93,GoeGra94,Wis96,GamWis01,WisDio01,GarZol00}. 
The form of this trajectory can be either jump-like in nature or
diffusive depending on how we choose to measure the system; that
is, the arrangement of the measuring apparatus. In this paper we
review quantum trajectory theory for partially observed systems by
presenting a simple model: A three level atom that emits into two
separate environments, only one of which is accessible to our detectors. Such partially
observed systems cannot be described by a stochastic \sch~equation (SSE)
\cite{Car93a, GarParZol92,MolCasDal93}, but rather requires a more general form
of a quantum trajectory that has been called a stochastic master equation (SME)
\cite{WisMil93a}. This is an instance of the fact that the  most general
form of quantum measurement theory requires the full Kraus
representation of operations \cite{Kra83,BraKha92}, rather than just
measurement operators
\cite{BraKha92}.

It is also well known that if we have a classical system and we
make measurements on it with a measurement apparatus that has
associated with it a Gaussian noise, then the evolution of this
classical state in the continuous-in-time limit obeys a
Kushner-Stratonovich equation (KSE) \cite{Mcg74}. To review these
dynamics for partially observed classical systems we present the
KSE for a system that experiences an `internal' unobservable white
noise process. That is, the evolution in the absence of the
measurements is given by a Fokker-plank equation \cite{Gar85}. This is the
classical analogue to the quantum master equation.

The new work in this paper is a simple numerical technique that
allows us to reduce the numerical resources required to calculate
the continuous-in-time trajectories. This method relies on the
implementation of linear or `ostensible' \cite{Wis96} measurement theory,
classical \cite{Mcg74} and quantum
\cite{GoeGra93,GoeGra94,Wis96,GamWis01}. For the classical case
our method reduces the problem from solving the KSE for the
probability distribution to simulating the ensemble average of two
coupled stochastic differential equations (SDE). For the quantum
case our method reduces the problem from solving a conditional
SME to simulating the ensemble average of a
SSE plus a c-number SDE. Thus in both
the classical and the quantum case, our method reduces the size of
the problem by a factor of $N,$ the number of basis states
required to represent the system.

Recently Brun and Goan \cite{Brun} have used a similar 
idea to investigate a partially observed quantum system.
However, since they did not use measurement theory with ostensible
probabilities,  their claim that they can generate a typical trajectory 
conditioned on some partial record ${\bf R}$  is not valid. 
This is demonstrated in detail in \ref{AppendixQuant}. (In their
method, the record ${\bf R}$ can only be generated randomly, and 
can be found only by doing the ensemble average over the fictitious noise, 
but that is not the issue of concern here.)

Finally, we combine these theories to consider the following case:
a quantum system is monitored continuously in time by a
classical system but we only have access to the results of
non-ideal measurements performed on the classical system. Note
that such joint systems have recently been studied by Warszawski
{\em et al} \cite{WarWisMab01,WarWis03a,WarGamWis04} and
Oxtoby {\em et al} \cite{Neil}. Warszawski {\em et al} considered
continuous-in-time monitoring of a quantum optical system with
realistic photodections while Oxtoby {\em et al} considered
continuous-in-time monitoring of a quantum solid-state system with
a quantum point contact. We show that our ostensible
numerical technique can be applied to these types of systems,
greatly simplifying the simulations.

The format of this paper is as follows. In Secs.~\ref{measQ} and
\ref{measC} we review quantum and classical measurement theory respectively.
This is essential as it allows us to define both the notations and
the physical insight that will be used throughout this paper. In
Secs.~\ref{quantum}, \ref{class}, and \ref{both} we investigate
the above mentioned quantum, classical and joint systems respectively, 
and present our ostensible numerical technique for each specific
case. Finally in Sec.~\ref{con} we conclude with a discussion.

\section{Quantum Measurement theory (QMT)}\label{measQ}

\subsection{General theory}

In quantum mechanics the most general way we can represent the
state of the system is via a state matrix $\rho(t)$. This is a
positive semi-definite operator that acts in the system Hilbert
space ${\cal H}_{\rm s}$. In this paper we take the view that this
represents our state of knowledge of the system. Taking this view
allows us to simply interpret the ``collapse of the
wavefunction'', upon measurement, as an update in the observer's
knowledge of the system \cite{CavFucSch02,Fuc02}. If we now assume
that we have a measurement apparatus that allows us to measure
observable $R$ of the system, then the conditional state $\rho_{r}(t')$ of the
system given result $r$ is determined by
\cite{Kra83}
\begin{equation}\label{QuantumUpdate}
\rho_{r}(t')=\frac{\hat{\cal O}_r(t',t)\rho(t)}
{P(r,t')},
\end{equation} where $P(r,t')$ is the probability of getting result
$r$ at time $t'=t+T$, where $T$ is the measurement duration time. Here
$\hat{\cal O}_r(t',t)$ is known as the operation of the
measurement and is a completely positive superoperator and for efficient measurements
can be defined by
\begin{equation}\label{OperationDef}
\hat{\cal O}_r(t',t)\rho(t)=\hat{\cal
J}[\hat{M}_{r}(T)]\rho(t)=\hat{M}_{r}(T)\rho(t)\hat{M}_{r}\dg(T),
\end{equation} where $\hat{M}_r(T)$ is called a measurement operator.
The probability of getting result $r$ is given by
\begin{eqnarray}\label{QuantumProb}
P(r,t')={\rm Tr}[\hat{\cal O}_r(t',t)\rho(t)]={\rm
Tr}[\hat{F}_r(T)\rho(t)],
\end{eqnarray} where the set $\{\hat{F}_r(T)=\hat{M}_r\dg(T)\hat{M}_r(T)\}$ is the positive
operator measure (POM) for observable $R$. By completeness, the
sum of all the POM elements satisfies
\begin{equation}\label{QuantumMeas}
\sum_r \hat{F}_r(T)=\hat{1}.
\end{equation}

So far we have only considered efficient, or purity-preserving measurements.
That is if $\rho(t)$ was initially $\ket{\psi(t)}\bra{\psi(t)}$
then the state after the measurement would also be of this form.
In a more general theory we must dispense with the measurement
operator $\hat{M}_r(T)$ and define the Kraus operator
$\hat{K}_{r,f}(T)$ \cite{Kra83}. This has the effect of changing
the definition of the operation of the measurement $\hat{\cal
O}_r(t',t)$ [\erf{OperationDef}] to
\begin{equation}\label{OperationDefComplete}
\hat{\cal O}_r(t',t)=\sum_f\hat{\cal
J}[\hat{K}_{r,f}(T)],
\end{equation} and the POM elements for this measurement are now given by
\begin{equation} \label{KrausEffec}
\hat{F}_{r}(T)=\sum_f\hat{K}_{r,f}\dg(T)\hat{K}_{r,f}(T).
\end{equation} Note $\hat{F}_{r}(T)$ still satisfies the completeness condition
[\erf{QuantumMeas}]. We can think of $f$ as labelling results of fictitious measurement.

If one is only interested in the average evolution of the system,
this can be found via
\begin{equation} \label{AverageO}
\rho(t')=\sum_r  P(r)\rho_r(t')=\hat{\cal O}(t',t)\rho(t),
\end{equation}
where $\hat{{\cal O}}(t',t)=\sum_r\hat{\cal O}_r(t',t)$ is the
non-selective operation.

\subsection{Quantum trajectory theory}

Quantum trajectory theory is simply quantum measurement theory
applied to a continuous in-time monitored system
\cite{Car93a,GarParZol92,WisMil93a,MolCasDal93,GoeGra93,GoeGra94,Wis96,GamWis01,WisDio01,GarZol00}.
In continuous monitoring, repeated measurements of duration $T=d
t$ are performed on the system. This results in the state being
conditioned on a record $\bfr$, which is a string containing the
results $r_k$ of each measurement from time 0 to $t$ but not
including time 0. Here the subscript $k$ refers to a measurement
completed at time $t_{k}=k d t$. From the record $\bfr$, the conditioned
state at time $t$ can be written as
\begin{equation}
\rho_{\bf R}(t)=\frac{\tilde\rho_{\bf R}(t)}{{P}(\bfr)},
\end{equation}
where $\tilde\rho_{\bf R}(t)$ is an unnormalized state defined by
\begin{equation}
\tilde\rho_{\bf R}(t)=\hat{\cal
O}_{r_{k}}(t_k,t_{k-1})\ldots\hat{\cal O}_{r_2}(t_2,t_1)\hat{\cal
O}_{r_1}(t_1,0)\rho(0).
\end{equation}
The probability of observing the record $\bfr$ is
\begin{equation}
P(\bfr)={\rm Tr}[\tilde\rho_{\bfr}(t)].
\end{equation}

If we now assume that the coupling between the apparatus (bath)
and the system is Markovian then the average state
\begin{eqnarray}
\rho(t)=\hat{\cal O}(t_k,t_{k-1})\ldots\hat{\cal
O}(t_2,t_1)\hat{\cal O}(t_1,0)\rho(0)
\end{eqnarray} is equivalent to the reduced state
\begin{equation}
\rho_{\rm red}(t)={\rm Tr}_{\rm bath}[\ket{\Psi(t)}\bra{\Psi(t)}],
\end{equation} which itself obeys the Master equation \cite{Lin76}
\begin{equation}\label{QuantumMaster}
\dot{\rho}(t)=\hat{\cal L}\rho(t)=-i[\hat{H},\rho(t)]+\sum_j\gamma_j\hat{\cal
D}[\hat{L}_j]\rho(t).
\end{equation} Here $\hat{\cal
D}[\hat{A}]$ is the superoperator defined by
\begin{equation}\label{DampSuper}
\hat{\cal
D}[\hat{A}]\rho=\hat{A}\rho\hat{A}\dg-\hat{A}\dg\hat{A}\rho/2-\rho\hat{A}\dg\hat{A}/2,
\end{equation} and represents dissipation of information about the system into the
baths.

\subsection{Fictitious quantum trajectories: the ostensible
numerical technique} \label{OstensibleQ}

If the system is only partly observed ($f$
in \erf{OperationDefComplete} represents the unobservable processes)
this state will be mixed. This is not a problem for simple systems but for a
large system a numerical simulation for $\rhoi(t)$ would be impractical. This brings
us to the goal of this section which is to demonstrate that $\rhoi(t)$ can be
numerically simulated by using SSEs, requiring less space to
store on a computer.

To do this we assume that a fictitious measurement with record
$\bff$ is actually made on the unobservable process. Then we can
expand $\rhoi(t)$ to
\begin{equation}\label{eq29}
\rhoi(t)=\sum_{\bff}\rhoif(t){ P}(\bff|\bfr),
\end{equation}
where
\begin{equation}\label{eq30}
\rhoif(t)=\psik\psib.
\end{equation} Here $\psik$ is a normalised state conditioned on both $\bff$ and
$\bfr$. In quantum trajectory theory this is defined as
\begin{equation}\label{eq31}
\psik=\frac{\upsik}{\rt{{P}(\bff,\bfr)}},
\end{equation} where
\begin{equation}\label{eq32}
\upsik=\hat{M}_{r_{k},f_{k}}(dt)....\hat{M}_{r_{1},f_{1}}(dt)\ket{\psi(0)}.
\end{equation} Here $r_{k}$ and $f_{k}$ are the results of the measurement
operator
\begin{equation}\label{eq33}
\hat{M}_{r_k,f_k}(dt)=\bra{r_k}\bra{f_k}\hat{U}(t_k,t_{k-1})\ket{0}\ket{0},
\end{equation} where $\ket{0}\ket{0}$ is the initial bath state.
This indicates that given that we have a real
record $\bfr$ we can calculate $\rhoi(t)$ from averaging over an
ensemble of pure states $\psik$. But as shown in
\ref{AppendixQuant} the fact that future real results are not
necessarily independent from the current fictitious results means that
we cannot generate single trajectories without knowing the full
solution. However by using quantum measurement theory with
ostensible distributions we can get around this problem.

Under ostensible quantum trajectory theory
\cite{GoeGra94,Wis96,GamWis01} we can define a state, $\lpsik$ as,
\begin{equation}\label{eq38}
\lpsik=\frac{\upsik}{\rt{\Lambda(\bff, \bfr)}},
\end{equation} where $\Lambda(\bff, \bfr)$ is an ostensible probability distribution.
This is simply a guessed distribution that only has the requirement that
it be a probability distribution and be non-zero when $P(\bff,
\bfr)$ is non-zero. Note this state is no longer normalized to one
and this is why we signify it with the bar. The true probability can be
related to the ostensible probability by
\begin{equation}\label{eq40}
{P}(\bfr,\bff)=\lpsib{{\bar{\psi}_{{\bf R, F
}}(t)}}\rangle\Lambda(\bff, \bfr),
\end{equation} which is a generalized Girsanov transformation
\cite{BelSta92,GoeGra94,Wis96,GamWis01,GatGis91}.

Going back to \erf{eq29} and using the above equations  we can
write $\rhoi(t)$ as
\begin{equation}\label{eq42}
\rhoi(t)=\frac{\sum_{\bff}\lpsik\lpsib\Lambda(\bff, \bfr)}{{
P}(\bfr)},
\end{equation}
where
\begin{equation}\label{eq43}
{P}(\bfr)=\sum_{\bff}\langle\bar{\psi}_{{\bf R}, {\bf F}}(t) \lpsik \Lambda(\bff,\bfr).
\end{equation}
Note that the sum containing $\Lambda(\bff, \bfr)$ in the above equations
simply represents the ensemble average over all possible
fictitious records. Thus we can rewrite \erf{eq42} as
\begin{equation}\label{MainEquation}
\rhoi(t)=\frac{{\rm E}_{\bf F} \Big{[}\lpsik\lpsib\Big{]}}
{{\rm E}_{\bf F}\Big{[}\langle\bar{\psi}_{{\bf R}, {\bf F}}(t) \lpsik\Big{]}}.
\end{equation}

\section{Classical Measurement theory (CMT)} \label{measC}
\subsection{General theory}

In this paper when considering what we call a classical system,
we are referring to a system that can be described by the
probability distribution $P(x,t)$ (i.e a vector of probabilities)
rather than a state matrix. That is, with respect to a fixed basis
$x$ the coherences (off diagonal elements) are always zero. If
we now measure observable $R$ of the system, then after a
measurement which yielded result $r$, the state of the system is
given by \cite{Bayes}
\begin{equation}\label{BayesTheorem}
P_r(x,t)=\frac{P(r,t|x,t)P(x,t)}{P(r,t)},
\end{equation}
where
\begin{equation}\label{Pr}
P(r,t)=\int dx P(r,t|x,t)P(x,t).
\end{equation} This is known as Bayes' theorem. Here $ P_r(x,t)\equiv P(x,t|r,t)$ is called
a conditional state and represents our new state of knowledge
given that we observed result $r$. Here we have only considered
minimally disturbing classical measurements. That is, there is no
back action acting on the system in the measurement process. To
generalize Bayes' theorem to deal with measurements which incur
back action we mathematically split the measurement into a two
stage process. The first is the Bayesian update, followed by a
second stage described by $ B_r(x',t'|x,t)$, the probability for
the measurement to cause the system to make a transition from $x$
at time $t$ to $x'$ at time $t'=t+T$, given the result $r$. Thus
for all $x'$, $x$ and $r$
\begin{eqnarray}\label{Bpropeties1}
B_r(x',t'|x,t)&\geq& 0,\\
\int dx' B_r(x',t'|x,t)&=& 1\label{Bpropeties}.
\end{eqnarray} Now by defining the operation
\begin{eqnarray}
{\cal O}_r(x',t'|x,t)&=& B_r(x',t'|x,t)P(r,t|x,t)
\end{eqnarray}
the conditional system state after the measurement becomes
\begin{equation}\label{GeneralBayessTheorem}
P_r(x',t')=\frac{ \int dx {\cal O}_r(x',t'|x,t)P(x,t)}{P(r,t')},
\end{equation} where
\begin{eqnarray}\label{Pr2}
P(r,t')&=&\int dx'\int dx  {\cal O}_r(x',t'|x,t) P(x,t).
\end{eqnarray} Using \erf{Bpropeties} this can be rewritten as
\begin{eqnarray}\label{Pr3}
P(r,t')&=&\int dx F_r(x,t) P(x,t),
\end{eqnarray}where $F_r(x,t)=P(r,t|x,t)$,
which by definition satisfies
\begin{equation}
\sum_r F_r(x,t)=1,
\end{equation} is the classical analogue of the POM element.
The average evolution of the system is given by
\begin{eqnarray}\label{Ave}
P(x',t')&=&\sum_{r}P_r(x',t')P(r,t')\nn\\&=&\int dx {\cal
O}(x',t'|x,t)P(x,t),
\end{eqnarray}
where ${\cal O}(x',t'|x,t)=\sum_r{\cal O}_r(x',t'|x,t)$ is the non-selective operation.

Note that for any $B_r(x',t'|x,t)$ that satisfies
\erfs{Bpropeties1}{Bpropeties} we can rewrite it as
\begin{equation}\label{Bansatz}
B_r(x',t'|x,t)=\sum_f \delta [x'-x_{r,f}(t')]P(f,t'|x,t;r,t),
\end{equation} where $x_{r,f}(t')$ is the new system configuration $x'$ at time $t'$
given the measurement result $r$ and
extra noise $f$ (the stochastic part of the back action). The parameter $f$ is
analogous to the fictitious measurement results in the quantum case. Thus the
operation for the measurement can be written as
\begin{eqnarray}\label{orf}
{\cal O}_r(x',t'|x,t)&=& \sum_f \delta [x'-x_{r,f}(t')]P(f,t';r,t|x,t),\nn\\ &=&\sum_f
{\cal J}_{r,f}(x',t'|x,t).
\end{eqnarray} This is the classical equivalent of \erf{OperationDefComplete}.

In the above we have purposely structured QMT and CMT so that the
theories appear to be similar and as a general rule we will push
this point of view throughout the rest of this paper. However, it
is important to point out the key differences between these
theories. In the quantum case we can always write the measurement
operator (or Kraus operator) as
$\hat{M}_r=\hat{U}_r\rt{\hat{F}_r}$ where $U_r$ is a unitary
operator. That is we can always interpret
a measurement as a two stage process, where $\rt{\hat{F}_r}$ is
responsible for the wavefunction collapse and the gain in
information by the observer and $\hat{U}_r$ is some extra
evolution that entails no information gain (as the entropy of
the system is not changed by this evolution). It simply adds
surplus back action to the system. In the classical case we can
also write the measurement as a two stage process. However, the
first process by definition has no back action; it is simply the
update in the observer's knowledge of the system. Furthermore the
second stage is not necessary unitary evolution (and as such can
change the entropy of the system). Thus back action in the quantum
and classical case are physically different processes and one can
not separate all the back action in the quantum case from the
observer's information gain. Mathematically speaking, the
difference arises from the fact that a quantum state is
represented by a positive {\em matrix}, the state matrix, while a
classical state is represented by a positive {\em vector}, the vector of
probabilities.

\subsection{Classical trajectory theory}

To achieve continuous-in-time measurements theory for a classical
system we simply let the measurement time tend to $dt$ and extend
the number of consecutive measurements to $t/dt$. Then the state
of the classical system given the measurement record $\bfr$ is
\begin{equation}
P_{\bfr}(x,t)=\frac{\tilde{P}_{\bfr}(x,t)}{{P}(\bfr)},
\end{equation}
where $\tilde{P}_{\bfr}(x,t)$ is an unnormalized state defined by
\begin{eqnarray}
&&\hspace{-.8cm} \tilde{P}_{\bfr}(x,t)=\int dx_{k-1}...\int
dx_1\int dx_0 \nl\times{\cal O}_{r_{k}}(x,t |x_{k-1},t_{k-1})
\ldots {\cal O}_{r_{2}}(x_2,t_2 |x_1,t_1) \nl\times{\cal
O}_{r_{1}}(x_1,t_1|x_0,0)P(x_0,0).
\end{eqnarray}
The probability of observing the record $\bfr$ is
\begin{equation}
P(\bfr)=\int dx \tilde{P}_{\bfr}(x,t).
\end{equation}

If we now assume that the noise added by the measurement apparatus
is white, and the form of the back action is independent of the
results ${\bf R}$, then the unconditional state
\begin{eqnarray}
&&\hspace{-.8cm}  {P}(x,t)= \int dx_{k-1}...\int dx_1\int dx_0
\nl\times{\cal O}(x,t |x_{k-1},t_{k-1})
\ldots {\cal O}(x_2,t_2 |x_1,t_1) \nl\times{\cal
O}(x_1,t_1|x_0,0)P(x_0,0)
\end{eqnarray}
is  the solution of the Fokker Plank Equation
\cite{Gar85}
\begin{equation}\label{Eq.FPE2}
\partial_t
P(x,t)=-\partial_{x}[A(x,t)
P(x,t)]+\smallfrac{1}{2}\partial^2_{x}[
D^2(x,t)P(x,t)],
\end{equation} where $A(x,t)$ determines the amount of drift and $D(x,t)$ determines the amount of
diffusion.

\subsection{Fictitious classical trajectories: The ostensible numerical
technique}\label{sec.fitcla}

The basic principle behind this technique is that we assume that
the unobservable process, $\bff$, that generates the back action part of the measurement is
fictitiously simulated. To be more
specific we can define
\begin{eqnarray}\label{ptilde}
&&\hspace{-.8cm} \tilde{P}_{\bfrf}(x,t)=\int dx_{k-1}...\int
dx_1\int dx_0 \nl\times{\cal J}_{r_{k},f_{k}}(x,t
|x_{k-1},t_{k-1})
\ldots {\cal J}_{r_{2},f_{2}}(x_2,t_2 |x_1,t_1) \nl\times{\cal
J}_{r_{1},f_{1}}(x_1,t_1|x_0,0)P(x_0,0).
\end{eqnarray}
where ${\cal J}_{r,f}(x',t'|x,t)$ is defined implicitly in
\erf{orf}. From this the conditional state, $P_{\bfr}(x,t)$, is
given by
\begin{equation}\label{conclass}
{P}_{\bfr}(x,t)=
\sum_{\bff}{P}_{\bfrf}(x,t){{P}({\bff|\bfr})},
\end{equation} where
\begin{equation}
{P}_{\bfrf}(x,t)=\frac{\tilde{P}_{\bfrf}(x,t)}{{P}({\bfr,\bff})}.
\end{equation} But as in the quantum case this cannot be directly
calculated and as a result we must use an ostensible theory. We
define the ostensible state by
\begin{equation}\label{pbar}
\bar{P}_{\bfrf}(x,t)=\frac{\tilde{P}_{\bfrf}(x,t)}{{\Lambda}({\bfr,\bff})},
\end{equation} and the true probability can be related to the
ostensible by
\begin{equation}
{{P}({\bfr,\bff})}=\int dx
\bar{P}_{\bfrf}(x,t){{\Lambda}({\bfr,\bff})},
\end{equation} the classical Girsanov transformation.

Using the above we can rewrite \erf{conclass} as
\begin{equation}\label{Pxgrvialin}
{P}_{\bfr}(x,t)=
\frac{\sum_{\bff}\bar{P}_{\bfrf}(x,t){{\Lambda}({\bff,\bfr})}}{P(\bfr)},
\end{equation} where
\begin{equation}
{P(\bfr)}=\sum_{\bff}\int dx
\bar{P}_{\bfrf}(x,t){{\Lambda}({\bff,\bfr})}.
\end{equation} As in the quantum case we can rewrite \erf{Pxgrvialin} as
\begin{equation}\label{EPxgrvialin}
{P}_{\bfr}(x,t)=
\frac{{\rm E}_{\bf F}\Big{[}\bar{P}_{\bfrf}(x,t)\Big{]}}{{\rm E}_{\bf F}\Big{[}\int dx\bar{P}_{\bfrf}(x,t)\Big{]}
},
\end{equation}
where $\bar{P}_{\bfr,\bff}(x,t)$ is an unnormalized pure classical
state. That is, it is of the form
$\bar{P}(x,t)=p_{\bfrf}\delta[x-x_{\bfrf}(t)]$, where $p_{\bfrf}$
is the norm of the ostensible state. To show this we consider a
system initially in the state $\bar{P}(x,0)=p\delta(x-x_0)$ then
by using \erfs{pbar}{ptilde} with ${\cal
J}_{r_{1},f_{1}}(x',t_1|x,0)$ defined implicitly in \erf{orf} we can rewrite $\bar{P}_{r_1,f_1}(x',t_1)$ as
\begin{equation}
\bar{P}_{r_1,f_1}(x',t_1)=p_{r_1,f_1}(t_1)\delta
[x'-x_{f_1,r_1}(t_1)],
\end{equation} which is still of the $\delta$-function form. Here 
$p_{r_1,f_1}(t_1)$ is given by
\begin{equation}\label{pdifff}
p_{r_1,f_1}(t_1) = P(f_1,t_1;r_1,0|x_0,0)p(0)/\Lambda({r_1,f_1}), 
\end{equation} and $x_{f_1,r_1}(t_1)$ is determined by the underlying
dynamics.  That is, we can simulate the distribution by solving
the two coupled SDEs, $\dot{x}_{\bf R, F}(t)$ and $\dot{{p}}_{\bf
R, F}(t)$.

\section{A Quantum system with an unobserved
process}\label{quantum}

\begin{figure}\begin{center}
\includegraphics[width=0.25\textwidth]{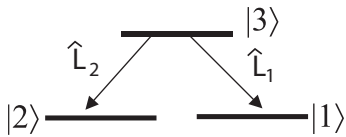}
\caption{\label{fig1} A simple system (a three level atom) which
has two outputs due to the to lowering operators $\hat{L}_1$ and
$\hat{L}_2$.}\end{center}
\end{figure}

To illustrate a quantum system where a complete measurement can
not be performed, due to some physical constraint, the system in
Fig.~\ref{fig1} was considered. This system is a three level atom
with lowering operators $\hat{L}_1=\ket{1}\bra{3}$ and
$\hat{L}_2=\ket{2}\bra{3}$, and decay rates $\gamma_{1}$ and
$\gamma_{2}$ respectively.

\subsection{Master equation}

\begin{figure}[t]\begin{center}
\includegraphics[width=0.45\textwidth]{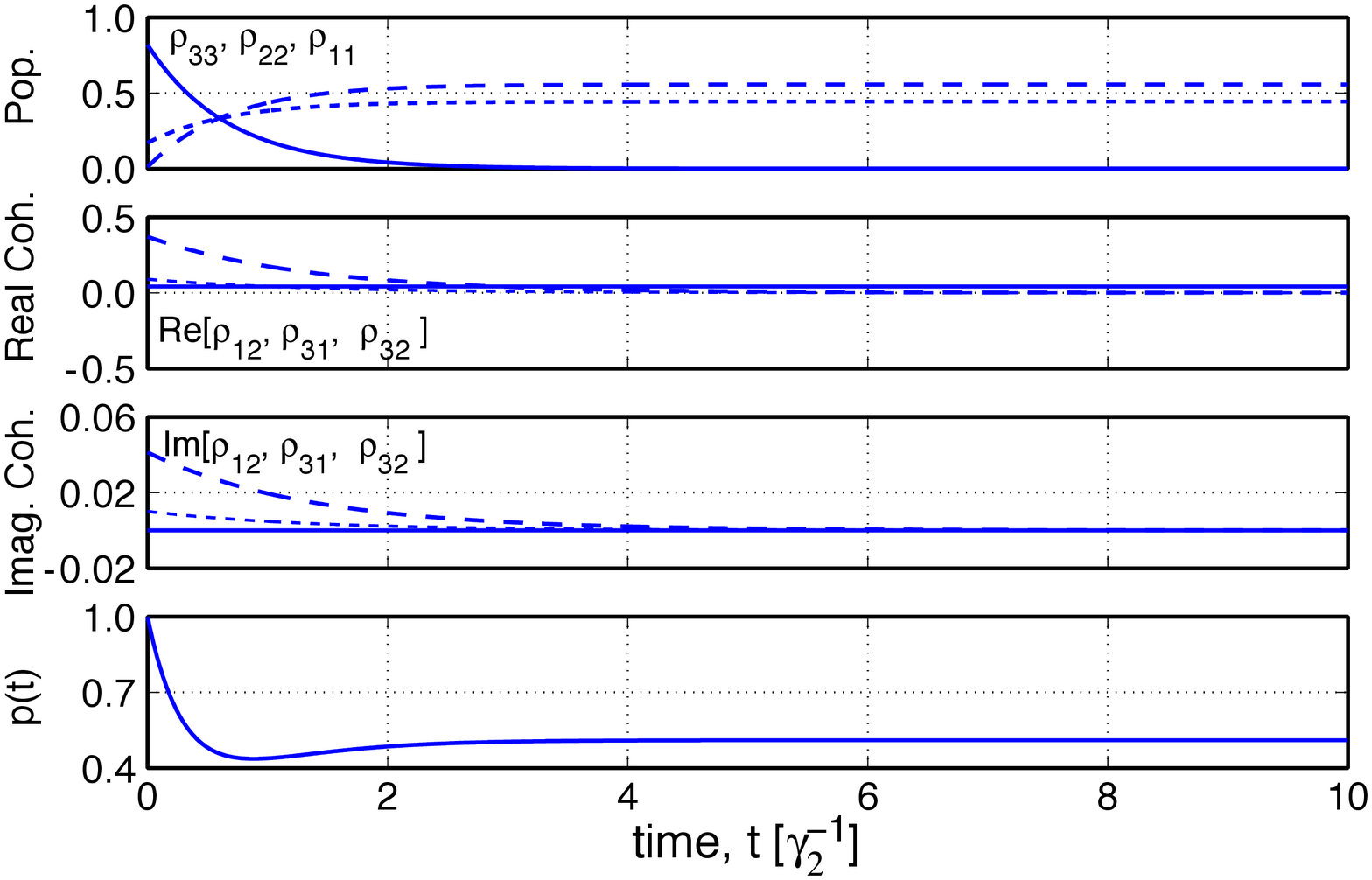}
\vspace{0.2cm} \caption{\label{fig2} The solution to the master
equation. The first subplot shows $\rho_{33}(t)$ (solid line),
$\rho_{22}(t)$ (dashed line) and $\rho_{11}(t)$ (dotted line). The
second and third subplot show the real and imaginary parts
respectively of $\rho_{12}(t)$ (solid line), $\rho_{31}(t)$
(dashed line) and $\rho_{32}(t)$ (dotted line). The fourth subplot
illustrates the purity of this state. This is all for the initial
condition
$\ket{\psi(0)}=0.4123\ket{1}+0.1\ket{2}+(0.9+0.1i)\ket{3}$ and
$\gamma_{1}=0.5$ and $\gamma_{2}=1$.}\end{center}
\end{figure}

With no external driving [$\hat{H}=0$ in \erf{QuantumMaster}], the
solution of the master equation can be determined analytically. To
illustrate a non-trivial solution we calculated this solution for
the initial condition
$\ket{\psi(0)}=0.4123\ket{1}+0.1\ket{2}+(0.9+0.1i)\ket{3}$ and
coupling constants $\gamma_{1}=0.5$ and $\gamma_{2}=1$. This is
shown in Fig.~\ref{fig2}. In this figure it is observed that as
time goes on, the state becomes mixed. This is seen as the purity
$p(t)={\rm Tr}[\rho^2(t)]$ of the state decays (although not monotonically)
as time increases. This figure also shows that the state becomes a mixture of the two
ground states, with the ground state associated with the larger
coupling constant being weighted more heavily, even though it
started with less weight.

\subsection{Conditional evolution: The quantum trajectory} \label{Incomplete}

\begin{figure}\begin{center}
\includegraphics[width=0.4\textwidth]{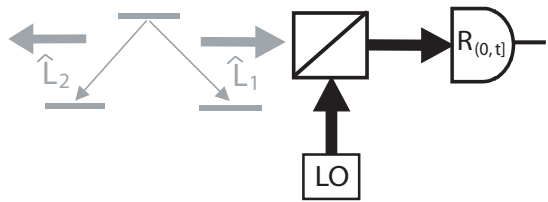}
\vspace{0.2cm} \caption{\label{fig3} A schematic representing
homodyne measurement of one of the outputs of the three level
atom. In an ordinary homodyne measurement the signal is coupled to
a classical local oscillator (LO) via a low reflective beam
splitter and then detected using a photoreceiver.}\end{center}
\end{figure}

In this section we consider the trajectory $\rho_{\bfr}(t)$ which
occurs when output $\hat{L}_1$ is monitored using homodyne-$x$
detection and output $\hat{L}_2$ is un-monitored. A schematic of
this measurement process is shown in Fig.~\ref{fig3}. Because this
arrangement is an inefficient measurement we have to use the
operation defined  in \erf{OperationDefComplete}. To determine the
Kraus operators we need to present the underlying dynamics in more
detail. For the interaction of this system with a Markovian bath
(and under the rotating wave approximation and in the interaction
frame) the total Hamiltonian is
\begin{eqnarray}\label{eq7}
H(t)&=&i\hbar\rt{\gamma_{1}}\int
\delta(t-t')[\hat{L}_1\hat{b}_r\dg(t')-\hat{L}_1\dg\hat{b}_r(t')]dt'
\nl+i\hbar\rt{\gamma_{2}}\int
\delta(t-t')[\hat{L}_2\hat{b}_f\dg(t')-\hat{L}_2\dg\hat{b}_f(t')]dt'.\nn\\
\end{eqnarray} Here $\hat{b}_r(t)$ and $\hat{b}_f(t)$ are the
temporal-mode annihilation operators for the detected
($\hat{b}_r$) and non-detected ($\hat{b}_f$) fields (baths). Since
these fields are Markovian there will be a commutator relationship
for the field of the following form
\begin{equation}\label{eq8}
[{\hat{b}_{i}(t),\hat{b}_{j}\dg(s)}]=\delta(t-s)\delta_{i,j},
\end{equation} where $i$, $j$ denotes either of the two baths.
This indicates that the field operators are gaussian white noise
operators. Thus they obey \ito calculus and the infinitesimal
evolution operator is \cite{GarParZol92,GarZol00}
\begin{eqnarray}\label{U}
\hat{U}(t+dt,t)&=&\exp\Big{\{}\rt{\gamma_{1}}[\hat{L}_1d\hat{B}_r\dg(t)-\hat{L}_1\dg
d\hat{B}_r(t)]\nl +\rt{\gamma_{2}}
[\hat{L}_2d\hat{B}_f\dg(t)-\hat{L}_2\dg d\hat{B}_f(t)]\Big{\}},
\end{eqnarray} where $d\hat{B}_i$ satisfies the commutator
relation
\begin{equation}
[d\hat{B}(t)_{i},d\hat{B}_{j}\dg(t)]=dt\delta_{i,j}.
\end{equation} Thus $\hat{U}(t+dt,t)$ is an operator acting in the Hilbert
space ${\cal H}_{s}\otimes {\cal H}_{r}\otimes {\cal H}_{f}$,
where ${\cal H}_{s}$, ${\cal H}_{r}$ and ${\cal H}_{f}$ are the
Hilbert spaces for the system, detected field and non detected
field respectively.

Now, given that a projective measurement is made on bath field
$\hat{b}_r(t)$ and bath field $\hat{b}_f(t)$ is completely
unobserved the state of the system after this measurement (time
$dt$ later) is given by \erfs{QuantumUpdate}{OperationDefComplete}
with $T=dt$, and the Kraus operator is
\begin{equation}\label{Kraus2}
\hat{K}_{r,f}(dt)=\bra{f}_f\bra{r}_r\hat{U}(t+dt,t)\ket{0}_{r}\ket{0}_f.
\end{equation} Here $\{\ket{r}_r\}$ is the set of orthogonal states the bath
is projected into, while $\{\ket{f}_f\}$ is any arbitrary
orthogonal basis set. For a homodyne-$x$ measurement of bath
$\hat{b}_r(t)$ the set $\{\ket{r}_r\}$ corresponds to the eigenset
of the operator $d\hat{B}_{r}(t)+d\hat{B}_{r}\dg(t)$
\cite{GoeGra94} and the results $r$ are the corresponding
eigenvalues. Note we have assumed that initially the baths, for
all the temporal-modes, are in the vacuum state.

After some simple rearrangement and using $(rdt)^2=dt$, the POM
elements for this measurement are of the form
\begin{equation}
\hat{F}_{r}(dt)=|\bra{r}{0}\rangle_{r}|^2[1+\rt{\gamma_{1}}r(t+dt)dt
\hat{x}_1],
\end{equation} where $\hat{x}_1=\hat{L}_1+\hat{L}_1\dg$. Thus
\begin{eqnarray}\label{eq20}
{ P}(r,t+dt)&=&|\bra{r}{0}\rangle|^2[1+rdt\rt{\gamma_{1}}\langle \hat{x}_1\rangle_t],
\end{eqnarray} where $\langle \hat{x}_1\rangle_t={\rm Tr}[\hat{x}_1\rho(t)]$.
Using the fact that $\ket{r}$ is a temporal-quadrature state,
\begin{equation}\label{eq22}
|\bra{r}{0}\rangle_{r}|^2=\rt{\frac{dt}{2\pi}}\exp\Big{(}-\frac{r^{2}}{2/dt}\Big{)},
\end{equation}  we can rearrange this to
\begin{equation}\label{eq23}
{ P}(r,t+dt)=\rt{\frac{dt}{2\pi}}\exp\Big{[}-\frac{[r-\rt{\gamma_{1}}\langle \hat{x}_1\rangle_t]^{2}}{2/dt}\Big{]}.
\end{equation} This implies that the random variable associated with this distribution, $r(t+dt)dt$, is a gaussian
random variable (GRV) of mean $\rt{\gamma_{1}}\langle
\hat{x}_1\rangle_t dt$ and variance $dt$. That is,
\begin{equation}\label{eq24}
r(t+dt) dt=dW(t)+dt\rt{\gamma_{1}}\langle \hat{x}_1\rangle_t,
\end{equation} where $dW(t)$ is a Wiener increment \cite{Gar85}.

Using the above and \erfs{QuantumUpdate}{OperationDefComplete} the
stochastic master equation for this system is
\begin{eqnarray}\label{eq16}
d\rho_{\bf R}(t+dt)&&=dt\Big{(}\gamma_{2}{\cal D}[\hat{L}_2]
+\gamma_{1}{\cal D}[\hat{L}_1]
\nl+dW(t)\rt{\gamma_{1}}{\cal
H}[\hat{L}_1]/dt\Big{)}\rho_{\bf R}(t),\nn\\
\end{eqnarray} where ${\cal H}[\hat{A}]$ is the superoperator
\begin{equation}\label{eq17}
{\cal H}[\hat{A}]\rho=\hat{A}\rho+\rho\hat{A}\dg -{\rm
Tr}[\hat{A}\rho+\rho\hat{A}\dg]\rho.
\end{equation}

\begin{figure}[t]\begin{center}
\includegraphics[width=0.45\textwidth]{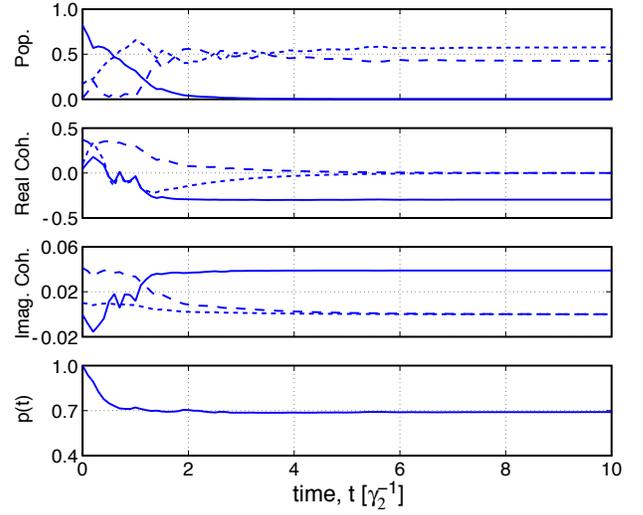}
\vspace{0.2cm} \caption{\label{fig4} The solution to $\rhoi$
written in matrix elements. The first subplot shows $\rho_{33,{\bf
R}}(t)$ (solid line), $\rho_{22,{\bf R}}(t)$ (dashed line),
$\rho_{11,{\bf R}}(t)$ (dotted line). The second and third subplot
show the real and imaginary parts respectively of $\rho_{12,{\bf
R}}(t)$ (solid line), $\rho_{31,{\bf R}}(t)$ (dashed line) and
$\rho_{32,{\bf R}}(t)$ (dotted line). The fourth subplot
illustrates the purity. We have used the same parameters as in
Fig. \ref{fig2}}\end{center}
\end{figure}

To illustrate an example quantum trajectory, \erf{eq16} was solved
for a randomly chosen record $\bfr$ and the same parameters used
in Fig. $\ref{fig2}$. This is shown in Fig. \ref{fig4}. It is
observed that this state evolution is stochastic in time and
becomes mixed (but not as mixed as the average evolution). It is
interesting to note that by performing this measurement the
coherence $\rho_{12, {\bf R}}(t)$, which was a constant of motion
for the average state, becomes comparable to the other coherence
and does not decay with time.

\subsection{The ostensible numerical technique}

In Sec.~\ref{OstensibleQ} we observed that the conditional
evolution of a partly monitored system could be simulated by
assuming that fictitious measurements are made on the unobservable
process.
For this system we assume that a fictitious homodyne-$x$ measurement is made on
output $\hat{L}_2$. Note we could have chosen any unraveling for $\bff$.

To determine the SSE for the ostensible state $\lpsik$ [the state
which we substitute into \erf{MainEquation} to determine the
actual conditional evolution] we have to derive the measurement
operator for the combined real and fictitious measurements, as well
as make a convenient choice for $\Lambda(\bff, \bfr)$. Using
\erf{eq33} and the fact that we are performing homodyne-$x$
measurements the measurement operator is
\begin{eqnarray}\label{eq34}
\hat{M}_{r,f}(dt)&=&\bra{f}0\rangle\bra{r}0\rangle\Big{(} 1+\rt{\gamma_{1}}rdt\hat{L}_1+\rt{\gamma_{2}}fdt\hat{L}_2\nl\hspace{-1cm}-\gamma_{1}dt
\hat{L}_1\dg\hat{L}_1/2-\gamma_{2}dt
\hat{L}_2\dg\hat{L}_2/2\Big{)},
\end{eqnarray} where the bath states $\ket{f}$ and $\ket{r}$ are
temporal quadrature states acting in Hilbert spaces ${\cal H}_f$
and ${\cal H}_r$ respectively. To derive this we have expanded \erf{U} to first order in $dt$ and used the fact that
$(fdt)^2=(rdt)^2=dt$. Since the real distribution is Gaussian
(with a variance $1/dt$) a convenient choice for $\Lambda(\bff,
\bfr)$ is $\Lambda(\bff)\Lambda(\bfr)$ where
$\Lambda(\bff)=\Lambda(f_k)\ldots\Lambda(f_1)$ and
$\Lambda(\bfr)=\Lambda(r_k)\ldots\Lambda(r_1)$ with
\begin{eqnarray}\label{o1}
\Lambda(r) &=& \rt{\frac{dt}{2\pi}}\exp\Big{[}-\frac{(r-\lambda)^{2}}{2/dt}\Big{]} \\
\Lambda(f) &=&
\rt{\frac{dt}{2\pi}}\exp\Big{[}-\frac{(f-\mu)^2}{2/dt}\Big{]}\label{ostenFit}.
\end{eqnarray} Here $\lambda$ and $\mu$ are arbitrary parameters.
With these ostensible distributions, \erf{eq34}, and \erf{eq38},
the ostensible SSE is
\begin{eqnarray}\label{eq50}
d\ket{\bar{\psi}_{\bf R, F}(t)}&=&dt
\Big{(}[r-\lambda](\rt{\gamma_{1}}\hat{L}_1-\lambda/2)+ [f-\mu]\nl\times(\rt{\gamma_{1}}\hat{L}_2-\mu/2)
-\smallfrac{1}{2} [\gamma_{1}\hat{L}_1\dg\hat{L}_1+\gamma_{2}\hat{L}_2\dg\hat{L}_2
\nl-\rt{\gamma_{1}}\lambda\hat{L}_1-\rt{\gamma_{2}}\mu\hat{L}_2+\lambda^2/4+\mu^2/4]
\Big{)}
\nl\times
\ket{\bar{\psi}_{\bf R, F}(t)}.
\end{eqnarray}

Now since we are interested in calculating $\rhoi(t)$ based on an
assumed known real record $\bfr$, we can rewrite \erf{eq50} as
\begin{eqnarray}\label{eq52}
dc_{1}&=&
c_{3}[\rt{\gamma_{1}}(r-\lambda)dt+dt\lambda/2]-c_1[\rt{\gamma_2}
d{\cal
W}\mu\nl+\rt{\gamma_1}(r-\lambda)dt\lambda+dt\lambda^2/4+dt\mu^2/4]/2,\\
dc_{2}&=&c_{3}[\rt{\gamma_{2}}d{\cal
W}+dt\mu/2]-c_2[\rt{\gamma_2} d{\cal
W}\mu\nl+\rt{\gamma_1}(r-\lambda)dt\lambda+dt\lambda^2/4+dt\mu^2/4]/2\\
dc_{3}&=&c_3[-\gamma dt+\rt{\gamma_2} d{\cal
W}\mu+\rt{\gamma_1}(r-\lambda)dt\lambda\nl+dt\lambda^2/4+dt\mu^2/4]/2,
\end{eqnarray} where $\gamma=\gamma_1+\gamma_2$. Here we have used the identity
\begin{equation}\label{eq47}
\ket{\bar{\psi}(t)}=c_{1}\ket{1}+c_{2}\ket{2}+c_{3}\ket{3},
\end{equation} and replaced
$fdt$ with $d{\cal W}(t)+\mu dt$, where $d{\cal W}(t)$
is a Wiener increment.

\begin{figure}[t]\begin{center}
\includegraphics[width=0.45\textwidth]{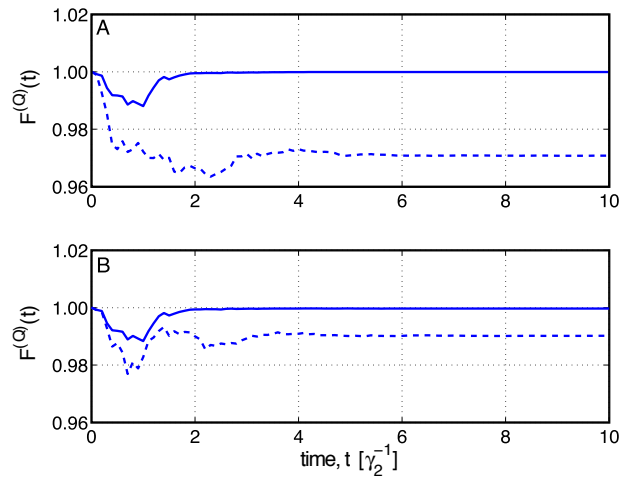}
\vspace{0.2cm} \caption{\label{felquantf} This figure shows the
fidelity between the actual $\rho_{\bfr}(t)$ and our ensemble
method for ensembles sizes 10 (dotted) and 1000 (solid). Part $A$
corresponds to a linear ostensible distribution while part $B$
refers to the non-linear ostensible distribution. The same
parameters were used as in Fig. \ref{fig2}.}\end{center}
\end{figure}

To illustrate the convergence of our method the ensemble average
of the above ostensible SSE for $\lambda=\mu=0$ was calculated for
$n=10$ and $n=1000$. To quantify how closely the ensemble method
reproduces $\rho_{\bfr}(t)$ we used the fidelity measure, which
for two different quantum states is defined as
\begin{equation}\label{felquant}
F^{\rm (Q)}(t)={\rm Tr}[\rt{\rt{\rho_1(t)}\rho_2(t)\rt{\rho_1(t)}}].
\end{equation} Note this measure ranges from 0 to 1 with 0
indicating two orthogonal states and 1 indicating the same state.
The result of this measure for the actual $\rho_{\bfr}(t)$ and the
ensemble version are shown in part A of figure \ref{felquantf}.
Here we see that for larger ensemble size the fidelity is closer
to one, indicating that as we increase the ensemble size our
ostensible method approaches the actual $\rho_{\bfr}(t)$.

To illustrate the effect of choosing different ostensible
distributions we considered the case when $\lambda=0$ and
\begin{equation}\label{eq422}
\mu=\rt{\gamma_2}\frac{\bra{\bar{\psi}_{\bf R,
F}(t)}\hat{L}_2+\hat{L}_2\dg \ket{\bar{\psi}_{\bf R,
F}(t)}}{\bra{\bar{\psi}_{\bf R, F}(t)}\bar{\psi}_{\bf R,
F}(t)\rangle}.
\end{equation} That is, the ostensible probability for the $k^{th}$ fictitious
results is the true probability we would expect based on the past
real and fictitious results up to, but not including the time
$kdt$. The motivation for this choice is that with $\mu=0$, the improbable
trajectories, ones that tend towards being inconsistent with the full real
record, will have norms that are very small and as such have little
contribution to the ensemble average. By contrast, using \erf{eq422}, the improbable
trajectories are less likely to be generated, so avoiding useless simulations.
With this ostensible distribution
the fidelity measure was calculated for $n=10$ and $n=1000$. These
results are shown in part $B$ of figure \ref{felquantf}. Here we
see that for the smaller ensemble size the fidelity is closer to
one than that observed using the first ostensible case. This
indicates that the rate of convergence for this case is greater
than the $\lambda=\mu=0$ case.

\section{A Classical system with an internal unobserved process}\label{class}

In this section we consider continuous-in-time measurements with
Gaussian precision of a classical system driven by an unobservable
noise process. This for example could correspond to a measurement
of the voltage across a resistor that is driven by a noisy classical
current.

\subsection{The average evolution}

We restrict ourselves to unconditional state evolution
described by the Fokker
Plank equation \erf{Eq.FPE2}. This equation has as its solution a
distribution that diffuses and drifts though time. Using Eq.
(\ref{Ave}) and only considering one interval in time we can write
\begin{equation}
P(x',t+dt)=\int dx {\cal O}(x',t+dt|x,t) P(x,t),
\end{equation} which when compared to \erf{Eq.FPE2} implies that
RHS of the above equation equals
\begin{eqnarray}\label{Eq.FPE3}
&&\hspace{-.8cm} \int dx [1 -dt\partial_{x'}A(x,t)
+dt\partial^2_{{x'}}D^2(x,t)/2]\delta(x'-x) \nl\times P(x,t).
\end{eqnarray} By introducing an arbitrary Gaussian distribution $P(f,t+dt)$ with mean
$m(t)$ and variance $1/dt$, that is
\begin{equation}\label{RealFclass}
P(f,t+dt)=
\rt{\frac{dt}{2\pi}}\exp\Big{[}-\frac{[f-m(t)]^{2}}{2/dt}\Big{]},
\end{equation} \erf{Eq.FPE3} can be rewritten as
\begin{eqnarray}\label{Eq.FPE4}
&&\hspace{-.8cm} \int df P(f,t+dt) \int dx [1 -dt\partial_{x'}A(x,t)
-dt [f-m(t)]\nl\times \partial_{x'}D_f(x,t)+dt\partial^2_{{x'}}D^2(x,t)/2]\delta(x'-x)
P(x,t).\nn\\
\end{eqnarray} By using \ito calculus and a Taylor expansion this can be
rewritten as
\begin{eqnarray}\label{Eq.FPE6}
&&\hspace{-.8cm}  \int dx {\rm E}_f \Big{\{}\delta[x'-x-dt
A(x,t)- dt[f(t+dt)-m(t)] \nl\times D(x,t)]\Big{\}}P(x,t).
\end{eqnarray} 
where $f(t+dt)dt=m(t)dt +d {\cal W}(t)$. Thus
\begin{eqnarray}\label{OO}
{\cal O}(x',t+dt|x,t)={\rm E}_f
\Big{\{}\delta[x'-x_f(t+dt)]\Big{\}},
\end{eqnarray} where $x_f(t+dt)$ is determined by the following SDE
\begin{equation} \label{Eq.Linear1}
d x_{\bf F}(t)=dt
A[x_{\bf F}(t),t]+ dt[f(t+dt)-m(t)]D[x_{\bf F}(t),t].
\end{equation} Note here we have written the SDE for the complete
record ${\bf F}$.

\subsection{Conditional evolution: The Kushner-Stratonovich
equation}\label{KSEsec}

To derive the KSE we start by deriving ${\cal O}_r(x',t'|x,t)$ and
$P(r,t+dt)$. For the case when the classical measurement has a
back action that is independent of the result $r(t+dt)$, the
operation for the measurement is given by
\begin{equation} {\cal
O}_r(x',t'|x,t)={\cal
O}(x',t'|x,t)P(r,t|x,t),
\end{equation} where $ {\cal  O}(x',t'|x,t)=B(x',t'|x,t)$.
Thus to derive ${\cal O}_r(x',t'|x,t)$  we need only $P(r,t|x,t)$.
For a measurement that has a precision limited by Gaussian white
noise it follows that
\begin{equation}\label{Eq.PIgivenx}
P(r,t|x,t)=F_r(x,t)=\frac{\sqrt{dt}}{\sqrt{2\pi \beta}}\exp[-(r-x)^2
dt/2\beta],
\end{equation} where $\beta$ is a constant characterizing the classical measurement strength.

To find $P(r,t)$ we substitute \erf{Eq.PIgivenx} into \erf{Pr3}.
This gives
\begin{equation}\label{Eq.PI}
P(r,t+dt)=\int dx\frac{\sqrt{dt}}{\sqrt{2\pi
\beta }}\exp[-(r-x)^2 dt/2\beta]P(x,t).
\end{equation} After some simple stochastic algebra and using  $r^2=\beta/dt$  this
can be simplified to \cite{WarWis03a}
\begin{equation}\label{Eq.PI5}
P(r,t+dt)
=\frac{\sqrt{dt}}{\sqrt{2\pi
\beta}} \exp[-(r-\langle x\rangle_t)^2dt/2\beta ],
\end{equation} where for the classical system $\langle x\rangle_t=\int
x P(x,t)dx$. From \erf{Eq.PI5} the stochastic representation of
$r(t+dt)$ is a Gaussian random variable with mean $\langle
x\rangle_t$ and variance $\beta dt$. That is,
\begin{equation}\label{Eq.I}
r(t+dt)=\langle x\rangle_t+\sqrt{\beta}dW(t)/dt.
\end{equation}

With all the above information and \erf{GeneralBayessTheorem} the
conditional state at time $t'=t+dt$ is
\begin{eqnarray}
&& \hspace{-0.8cm} P_r(x',t+dt)=\int dx {\rm E}_f \Big{\{}\delta[x'-x_f(t+dt)]\Big{\}}
\Big{\{}1\nl+[x-\langle x\rangle_t]
[r-\langle x\rangle_t]dt/\beta\Big{\}}P(x,t).
\end{eqnarray}
Here we have expanded the exponentials in \erf{Eq.PI5} and
\erf{Eq.PIgivenx} to second order in $dt$ and used $r^2=\beta/dt$.
Taylor expanding the delta function and averaging over the
$f(t+dt)$ [using \erf{RealFclass}] for each step in time gives the
KSE
\begin{eqnarray} \label{KS}
P_{\bf R}(x,t+dt)&=&P_{\bf R}(x,t)+dt[x-\langle
x\rangle_t][r(t+dt)-\langle x\rangle_t]\nl\times P_{\bf
R}(x,t)/\beta -dt\partial_x [{A}({x},t)P_{\bf
R}(x,t)]\nl+\smallfrac{1}{2} dt\partial^2_x [D^2({x},t)P_{\bf
R}(x,t)]
\end{eqnarray} and $\langle x\rangle_t$ becomes $\int
x P_{\bf R}(x,t)dx$. In general to
solve this equation we need to solve for all $x$. For some
${A}({x},t)$ and $D(x,t)$ this can be a rather lengthy numerical
problem. In the following section we will present our ostensible
technique which allows us to reformulate the problem to solving
two coupled SDEs, at the cost of performing an ensemble average.

\subsection{The ostensible numerical
technique}\label{sec.linearclass}

As shown in Sec. \ref{sec.fitcla}, if we consider the
unobservable process ${\bf F}$ as actually occurring then we can
simulate the KSE by using \erf{EPxgrvialin}, and 
$\bar{P}_{\bfr,\bff}(x,t)$ is determined by solving two coupled
SDEs. For the case when the classical measurement has Gaussian
precision and the back action only depends on the white noise
process $f(t)$, we can rewrite $P(f,t';r,t|x,t)$ in \erf{pdifff} as
$P(f,t')P(r,t|x,t)$ where $P(f,t')$ is given by \erf{RealFclass}
and $P(r,t|x,t)$ is given by \erf{Eq.PIgivenx}. Thus $\dot{x}_{\bf
R,F}(t)$ becomes $\dot{x}_{\bf F}(t)$ and is given by
\erf{Eq.Linear1}. To find the differential equation for
$\dot{p}_{\bf R, F}(t)$ we need to assume a form for the
ostensible distribution
$\Lambda(f,r)$. We use $\Lambda(f,r)=\Lambda(f)\Lambda(r)$, where
\begin{eqnarray}\label{ostenRit}
\Lambda(r) &=& \rt{\frac{dt}{2\pi}}\exp\Big{[}-\frac{(r-\lambda)^{2}}{2\beta/dt}\Big{]}
\end{eqnarray} and $\Lambda(f)$ is given by \erf{ostenFit}. Extending \erf{pdifff} to continuous measurements gives
\begin{eqnarray}\label{Eq.Linear2}
d p_{\bf R, F}(t)&=&
dt[m(t)-\mu][f(t+dt)-\mu]p_{\bf R, F}(t)\nl+dt[x(t)-\lambda][r(t+dt)-\lambda]p_{\bf R,
F}(t)/\beta.\nn\\
\end{eqnarray}
Thus to determine $\bar{P}_{\bf R, F}(x,t)$ we only need to
simulate \erfs{Eq.Linear1}{Eq.Linear2} with ${\bf R}$ assumed
known and $f(t+dt)$ given by \erf{ostenFit}. $P_{\bf R}(x,t)$ is
then determined by \erf{Pxgrvialin}. Since the theory
requires $\bar{P}_{\bf R, F}(x,t)$ to be a delta function, one might
conclude that this method is only valid for initial conditions of
the form ${P}(x,0)\rho(0)=\delta(x-x_0)$. Infact, we are not limited to this case.
To consider other initial conditions we simply choose the initial
value $x_0$ in \erf{Eq.Linear1} from the distribution ${P}(x,0)$.

\subsection{A simple example}

\begin{figure}\begin{center}
\includegraphics[width=0.45\textwidth]{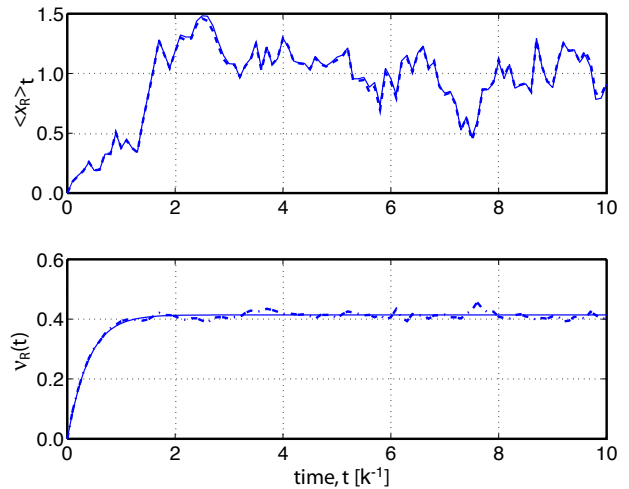}
\caption{\label{Fig.PxGivenIEquations} The mean and variance of
${P}_{\bfr}(x,t)$ when $\beta= 1$, $A=1-x$ and $D=1$ for both
${P}_{\bfr}(x,t)$ calculated exactly (solid) and via the linear
method for an ensemble size of 10 000 (dotted). }\end{center}
\end{figure}

To illustrate the classical theory we consider a Gaussian
measurement of a classical system that is driven by an an
unobservable white noise process with  $m(t)=0$ and drift and
diffusion functions given by \begin{eqnarray} \label{cond1}
A({x},t) &=& -kx+l, \\
D({x},t) &=& b.\label{cond2}
\end{eqnarray}  If this is the case then
$P_{\bfr}(x,t)$ has a Gaussian solution with a mean $\langle
x_{\bf R}\rangle_t$ and variance $\nu_{\bf R}(t)$ given by
\begin{eqnarray}
d\langle x_{\bf R}\rangle_{t}&=&dt\{\nu_{\bf R}(t)[r(t+dt)-\langle x_{\bf R}\rangle_t
]/\beta- k \langle x_{\bf R}\rangle_t\nl+l\},\label{Eq.bKS22}\\
\label{Eq.vKS22}
d\nu_{\bf R}(t)&=&dt[-\nu_{\bf R}^2(t)/\beta-2k\nu_{\bf R}(t)+b^2],
\end{eqnarray} and $r(t+dt)=\langle x_{\bf R}\rangle_t+dW(t)$. That is,
as time increases the measurement has
the effect of reducing the variance but the diffusive coefficient
$b$ causes this variance to increase. The mean, however contains
both the deterministic evolution and a random term due the
measurement. To illustrate this solution we have
simulated \erfs{Eq.bKS22}{Eq.vKS22} for the case when $ A(x,t) = 1-x$,
$D=1$ and $\beta=1$. The results of this simulation are shown in
Fig. \ref{Fig.PxGivenIEquations} as a solid line. Here we see that
the mean follows some stochastic path conditioned on the record
$\bfr$, while the variance follows a smooth function.

\begin{figure}\begin{center}
\includegraphics[width=0.45\textwidth]{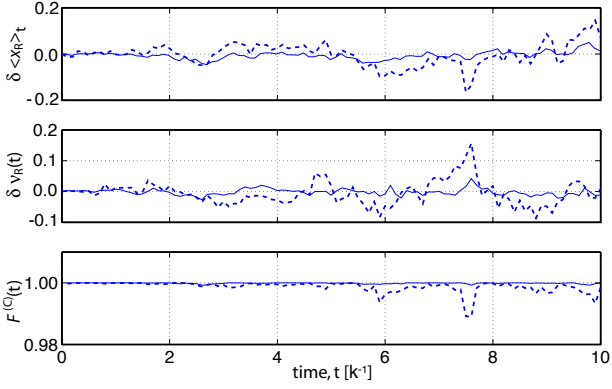}
\caption{\label{Fig.EPxGivenIF1000} The first and second plot show the difference between the
mean and variance of ${P}_{\bfr}(x,t)$ calculated by the linear
method and the know result for ensemble sizes 100 (dotted) and 10
000 (solid). The third plot shows the Fidelity between ${P}_{\bfr}(x,t)$ calculated by the linear
method and the know result for ensemble sizes 100 (dotted) and 10
000 (solid).  The parameters are the same as in Fig.
\ref{Fig.PxGivenIEquations}.}\end{center}
\end{figure}

To illustrate our ostensible method we use the above record and
solve numerically \erfs{Eq.Linear1}{Eq.Linear2} with
$\lambda=\mu=0$. The mean
and variance is then found via
\begin{eqnarray}
\langle x_{\bf R}\rangle_{t}&=&\frac{{\rm E}_{\bf F}\Big{[}{x}_{\bf
F}(t)p_{\bf R, F}(t)\Big{]}}{
{\rm E}_{\bf F}\Big{[}p_{\bf R, F}(t)\Big{]}} \\
\nu_{\bf R}(t)&=&\frac{{\rm E}_{\bf F}\Big{[}x_{\bf F}^2(t)p_{\bf R,
F}(t)\Big{]}}{
{\rm E}_{\bf F}\Big{[}p_{\bf R, F}(t)\Big{]}}-\langle x_{\bf R}\rangle^2_{t}\nn\\
\end{eqnarray}
where ${\rm E}_{\bf F}$ denotes an ensemble average over all
possible fictitious records. The numerical values for the mean and
variance are shown in Fig. \ref{Fig.PxGivenIEquations} (dotted)
for an ensemble size of 10 000. To get an indication of the
numerical error in the solution from our method, the difference from
the exact solution is shown in Fig. \ref{Fig.EPxGivenIF1000}. The
dotted line corresponds to an ensemble of 100 and the solid to one
of 10 000. Here we see that the ostensible method solution agrees well with
the exact solution and as we increase the ensemble size the
difference between these solutions decreases. To get a better
indication of how well our method reproduces the actual
${P}_{\bfr}(x,t)$, we also calculated the classical fidelity, which is defined by
\begin{equation} \label{felclass}
F^{(C)}(t)=\int dx \rt{P_1(x,t)}\rt{P_2(x,t)}.
\end{equation} This was calculated under the assumption that
the state calculated via the ostensible
method was also Gaussian. This is illustrated in Fig.
\ref{Fig.EPxGivenIF1000}, where we see that for the larger
ensemble the fidelity is very close to one, implying that the
distributions are almost identical.

\section{An unobservable quantum system driving a Classical
system}\label{both}

In this section we consider the following situation: a quantum
system is monitored continuously in time by a classical system. This
in turn is measured with Gaussian precision, and these are the only results
to which we have
access. This for example occurs when the signal
from the quantum system enters a detector with a bandwidth $B$,
resulting in the state of the detector being related to $\bff$ by
\cite{WarWis03a}
\begin{equation}\label{pollysyst}
x(t)=\int_{-\infty}^{t} ds B\exp[-B(t-s)]{f}(s).
\end{equation} Thus in a measurement that reveals $x(t)$ with perfect
precession [eg $F_r(x)=P(r,t|x,t)=\delta(r-x)$] we could determine $\bff$
(the quantum signal) by inverting the convolution in \erf{pollysyst}. But if this
measurement has Gaussian precision [\erf{Eq.PIgivenx}] then we
must treat the state of the detector as a classical probability
distribution and use a mixture of CMT and QMT to describe the
conditional state of the supersystem (classical and quantum
system). To denote the supersystem we use the
notation $\rho(x,t)$, where $x$ refers to the classical
configuration space and $\rho$ denotes an object acting on a
Hilbert space. This has the interpretation whereby $P(x,t)={\rm Tr}[\rho(x,t)]$ is the
(marginal) classical state and $\rho(t)=\int \rho(x,t)dx$ is the (reduced) quantum state.
For uncorrelated quantum and classical states, $\rho(x,t)=P(x,t)\rho(t)$.

\subsection{Conditional evolution} \label{ConSec}

We denote the state of the supersystem conditioned on the
classical result $r$ at time $t+dt$ as $\rho_{r}(x,t+dt)$. Assuming that the quantum
system is not affected by the classical system, this
can be expanded as
\begin{equation}\label{s}
\rho_{r}(x,t+dt)=\sum_{f}P_r(f,t+dt)P_{r,f}(x,t+dt)\rho_f(t+dt),
\end{equation} where $\rho_f(t+dt)$ is the state that an observer who had access to all the
quantum information would ascribe to the quantum system. That is,
$f(t+dt)$ can be regarded as really existing (with the collapse of the wavefunction
occurring at this level); it is just that the real observer does not
have access to this information. The state of knowledge of this real observer
is different from, but consistent with, that of the hypothetical
observer who has access to {\bf F}.

In terms of the operation of
the measurement, the conditional state can be written as
\begin{eqnarray}\label{ss}
\rho_{r}(x',t+dt)&=&\frac{\tilde{\rho}_{r}(x',t+dt)}{P(r,t+dt)},
\end{eqnarray} where
\begin{eqnarray}\label{sss}
\tilde{\rho}_{r}(x',t+dt)&=&\int dx
\sum_{f}  {\cal J}_{r,f}(x',t+dt|x,t)  \nl\times\hat{\cal O}_f(t+dt,t)
\rho(x,t)/P(f,t+dt) \hspace{.8cm}
\end{eqnarray} and
\begin{eqnarray}\label{ssss}
P(r,t+dt)&=&\int dx' {\rm Tr}\Big{[}\tilde{\rho}_{r}(x',t+dt) \Big{]}.
\end{eqnarray} The quantum part of the operation of measurement in defined by
\erf{OperationDefComplete} and the classical part is defined in
\erf{orf} with the replacement of $P(f,t';r,t|x,t)\rightarrow P(f,t')P(r,t|x,t)$ because in
this system the quantum signal does not depend on the classical state.

To illustrate the above we consider the case when we are
monitoring with Gaussian precision the classical system defined by
\erf{pollysyst} which is in turn monitoring the $x$
quadrature flux coming from a classically driven two level atom (TLA).
This is the same as the system considered in Ref \cite{WarWis03a}
and as such we will simply list the important equations. The
quantum part of operation is given by $\hat{\cal
O}_f(t+dt,t)=\hat{\cal J}[\hat{M}_f(dt)]$ where
\begin{equation}\label{m}
\hat{M}_f(dt)=\bra{f}0\rangle[1-dt(i\hat{H}-\rt{\gamma}f\hat{\sigma}+\gamma
\hat{\sigma}\dg\hat{\sigma}/2)].
\end{equation} The fictitious quantum signal statistic obeys
\begin{equation}
P(f,t+dt)=\int dx {\rm Tr}[\hat{\cal O}_f(t+dt,t)\rho(x,t)],
\end{equation} which for a homodyne-$x$ measurement
can be shown to be of the form displayed
in \erf{RealFclass} with $m(t)=\rt{\gamma}{\rm Tr} [(\hat
\sigma+\hat\sigma\dg)\rho(t)]$. Here $\hat{\sigma}$ is the
lowering operator for the TLA and $\gamma$ is the decay rate. Note
here we have assumed all the quantum signal is fed into the
classical system, if we wanted to simulate some inefficiency we
would simply use the Kraus represention, and for the case
where this inefficiency is a constant, $\eta$, we simply replace
$\sigma$ in the above equations by $\rt{\eta}\sigma$.

As shown in Sec. \ref{class} for a classical measurement with
Gaussian precision and a back action that does not depend on the
results of the measurement, the classical part of the operation is
\begin{equation}\label{p}
{\cal
J}_{r,f}(x',t+dt|x,t)=\delta[x'-x_f(t+dt)]P(f,t+dt)P(r,t|x,t),
\end{equation} where $P(r,t|x,t)$ is defined in \erf{Eq.PIgivenx}
and $x_f(t+dt)$ is given by \erf{Eq.Linear1}. For the system we
are considering, to find $A(x,t)$ and $D(x,t)$ we simply
differentiate \erf{pollysyst} and equate this with
\erf{Eq.Linear1}. Doing this gives
\begin{eqnarray}\label{Ad}
A(x,t)&=&-B x +B m(t),\\
D(x,t)&=&B. \label{Dd}
\end{eqnarray}

Combining the quantum and classical parts  of the operation and
using the same techniques as in Sec. \ref{KSEsec} allows us to
rewrite \erf{ss} for continuous-in-time measurements as
\begin{eqnarray}\label{supertrajextory}
d\rho_{\bf R}(x,t)&=&dt \Big{(} B\partial_{x}x
+\smallfrac{1}{2}B^2\partial^2_{x}
+\hat{\cal
L}\Big{)}\rho_{\bf R}(x,t) \nl+dt
\Big{(}\frac{[x-\langle x_{\bf R}\rangle_t][r(t+dt)-\langle x_{\bf R}\rangle_t]}{\beta}\Big{)}\rho_{\bf R}(x,t)
\nl-dt\rt{\gamma}\partial_x  B [\hat\sigma\rho_{\bf R}(x,t)+\rho_{\bf R}(x,t)\hat\sigma\dg],
\end{eqnarray} where $\langle x_{\bf R}\rangle_t=\int x{\rm Tr}[ \rho_{\bf R}(x,t)]dx $ and
\begin{equation}\label{rboth}
r(t+dt)dt= \langle x_{\bf R}\rangle_tdt+\rt{\beta}dW(t).
\end{equation} This equation (\ref{supertrajextory}) has been labeled the
Superoperator-Kushner-Stratonovich equation \cite{WarWis03a} and
represents the evolution of the combined supersystem. The first
line contains the free evolution for both the quantum and the
classical systems. For this quantum system
\begin{equation}
\hat{\cal L}[\hat\sigma]\rho=\frac{-i \Omega}{2}[\hat{\sigma}_x,\rho]
+\gamma\hat{\cal D}[\hat\sigma]\rho,
\end{equation} where $\Omega$ is the Rabi frequency and $\hat{\cal D}$ is the
damping superoperator and is
defined in \erf{DampSuper}. The second line of Eq.~(\ref{supertrajextory}) describes the gaining of
knowledge about the state of classical system via Gaussian
measurements. Lastly the third line describes the coupling of the
quantum and classical system.

For a TLA we can write
the state of the supersystem as
\begin{equation}\label{superrho}
\rho(x,t)=\smallfrac{1}{2}[P(x,t)\hat{1}+X(x,t)\hat{\sigma}_x+Y(x,t)\hat{\sigma}_y+Z(x,t)\hat{\sigma}_z].
\end{equation} Note that $P(x,t)$ is the marginal state of knowledge for the classical
system (found via tracing out the
quantum degrees of freedom).
Substituting this into \erf{supertrajextory} gives the following
four coupled partial differential equations
\begin{eqnarray}\label{classical}
\dot{P}_{\bf R}&=&[x-\langle x_{\bf R}\rangle_t][r-\langle
x_{\bf R}\rangle_t]P_{\bf R}/\beta+B\partial_x[x P_{\bf R}\nl-\rt{\gamma}X_{\bf R}]
+\smallfrac{1}{2}B^2\partial_x^2P_{\bf R}\\
\dot{X}_{\bf R}&=&[x-\langle x_{\bf R}\rangle_t][r-\langle
x_{\bf R}\rangle_t]X_{\bf R}/\beta+\smallfrac{1}{2}B^2\partial_x^2X_{\bf R}\nl+B\partial_x[x
X_{\bf R}-\rt{\gamma}P_{\bf R}-\rt{\gamma}Z_{\bf R}]
-\smallfrac{1}{2}\gamma X_{\bf R}\\
\dot{Y}_{\bf R}&=&[x-\langle x_{\bf R}\rangle_t][r-\langle
x_{\bf R}\rangle_t]Y_{\bf R}/\beta+B\partial_x[x Y_{\bf R}]\nl+\smallfrac{1}{2}
B^2\partial_x^2Y_{\bf R}-\Omega Z_{\bf R}-\smallfrac{1}{2}\gamma Y_{\bf R}\\
\dot{Z}_{\bf R}&=&[x-\langle x_{\bf R}\rangle_t][r-\langle
x_{\bf R}\rangle_t]Z_{\bf R}/\beta+B\partial_x[x Z_{\bf R}\nl+\rt{\gamma}X_{\bf R}
]+\smallfrac{1}{2}B^2\partial_x^2Z_{\bf R}+\Omega Y_{\bf R}-\gamma(P_{\bf R}+Z_{\bf R}).\nn
\\ \label{classical4}
\end{eqnarray} To determine the state of knowledge for the quantum
system we simply integrate out the classical degrees of freedom.

\begin{figure}\begin{center}
\includegraphics[width=0.45\textwidth]{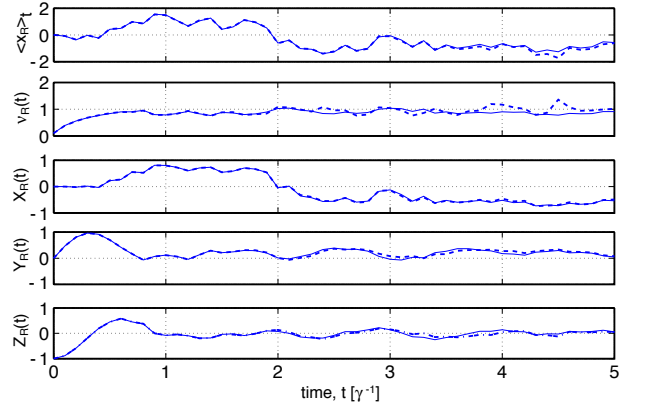}
\caption{\label{figsuperI} ${\rho}_{\bfr}(x,t)$ calculated via
numerical integration (solid) and via the ostensible method for an
ensemble size of 10 000 (dotted).  The parameters are $\beta=0.5$,
$B=2$, $\gamma=1$ and $\Omega=5$ and initial conditions
${\rho}(x,0)=P(x)\ket{g}\bra{g}$ where $P(x)$ is a Gaussian with
mean zero and variance 0.1. }\end{center}
\end{figure}

To illustrate a trajectory for this supersystem the following
parameters were used; $\beta=0.5$, $B=2$, $\gamma=1$ and $\Omega=5$.
The results are shown in Fig. \ref{figsuperI} (solid line) for
a randomly chosen record $\bfr$. This figure displays the mean and
the variance of the classical trajectory found via tracing over
the quantum degrees of freedom as well as the quantum state in
Bloch representation after we have integrated out the classical
degrees of freedom.

\subsection{Fictitious trajectories: The ostensible numerical technique}

In the above section we observed that to be able to calculate the
supersystem trajectory we needed to solve four coupled partial differential
functions (each involving derivatives with respect to a classical configuration coordinate $x$).
This is a rather lengthy calculation which for higher dimensional
($d$) quantum systems will require $d^2$ partial differential equations. Here we
present our linear method that allows us to reduce the problem to
$d+2$ couple differential equations. The
expense, again, is that an ensemble average must be performed.

To do this we simply note that we can define the following quantum
and classical states
\begin{equation}\label{linearquantum}
\bar\rho_f(t+dt)=\frac{\hat{\cal O}_f(t+dt,t)\rho(t)}{\bar{\Lambda}(f)}
\end{equation} and
\begin{equation}\label{linearclassical}
\bar P_{r,f}(x',t+dt)=\frac{\int dx \bar{\cal O}_{r,f}(x',t+dt|x,t)
P(x,t)}{\Lambda(f)\Lambda(r)},
\end{equation} where
\begin{equation}
\bar{\cal O}_{r,f}(x',t+dt|x,t)=\delta[x'-x_{f}(t+dt)]P(r,t|x,t)\bar{\Lambda}(f).
\end{equation} Note the bar above $\bar{\Lambda}(f)$ means that the ostensible distribution used to
scale the quantum state does not have to be the same as that used
to scale the classical state. Here for simplicity we
consider only the case when they are the same (as no numerical
advantage is gain by different choices). Using the above equations
we can rewrite \erfs{ss}{ssss} as
\begin{eqnarray}\label{li}
\rho_{r}(x',t+dt)&=&\frac{\sum_{f}\Lambda(f)\Lambda(r)\bar{P}_{r,f}(x',t+dt)\bar\rho_f(t+dt)}{P(r,t+dt)},\nn\\\\
P(r,t+dt)&=&\int dx \sum_f{\rm Tr}
[\Lambda(f)\Lambda(r)\bar{P}_{r,f}(x,t+dt)\nl\times\bar\rho_f(t+dt)].
\end{eqnarray} Thus to simulate $\rho_{\bf R}(x,t)$ we need only to calculate
$\bar{P}_{\bf R, F}(x,t)$ and $\bar\rho_{\bf F}(t)$ for a specific
record $\bfr$.

For the above TLA-classical detector system with $\Lambda(r)$ and
$\Lambda(f)$ defined by \erfs{ostenRit}{ostenFit} respectively,
$\bar P_{\bf R, F}(x',t)$ has a solution of the form $p_{\bf
R}(t)\delta[x'-x_{\bf R, F}(t)]$ where $x_{\bf F}(t)$ is given by
\begin{equation} \label{Eq.Linear1a}
d{x}_{\bf F}(t)=dt [-B x_{\bf F}(t) +B f(t+dt)]
\end{equation} and $p_{\bf R, F}(t)$ is given by
\begin{equation}\label{Eq.Linear3}
dp_{\bf R, F}(t)=dt[x_{\bf F}(t)-\lambda][r(t+dt)-\lambda]p_{\bf R, F}(t)/\beta,\hspace{1cm}.
\end{equation}
Thus we can rewrite \erf{li} as
\begin{equation}\label{liMain}
\rho_{\bf R}(x,t)=\frac{{\rm E}_{\bf F}\Big{[}\delta[x-x_{\bf F}(t)]p_{\bf R,F}(t)\bar\rho_{\bf F}(t)\Big{]}}
{{\rm E}_{\bf F}\Big{[}p_{\bf R,F}(t)\check{p}_{\bf F}(t) \Big{]}},
\end{equation} where $\check{p}_{\bf F}(t) ={\rm Tr}[\bar\rho_{\bf F}(t)]$.

To determine the evolution of the ostensible quantum state we
simply substitute the measurement operator defined in \erf{m} with
$\hat{H}=\Omega\hat{\sigma}_x/2$ and the ostensible distribution
${\Lambda}(f)$ into \erf{linearquantum}. Doing this gives
\begin{eqnarray}\label{linearquantumtraj}
d\bar\rho_{\bf F}(t)&=&dt\frac{-i\Omega}{2}[\hat{\sigma}_x,\rho_{\bf F}(t)]+dt\gamma\hat{\cal
D}[\hat{\sigma}]\rho_{\bf F}(t)+\nl dt [f(t+dt)-{\mu}][\rt{\gamma}\hat{\sigma}\rho_{\bf F}(t)
\nl+\rt{\gamma}\rho_{\bf F}(t)\hat{\sigma}\dg-{\mu}\rho_{\bf F}(t)].
\end{eqnarray} However since we have assumed that all the quantum
signal is fed into the detector the evolution of the ostensible quantum state can be
written as an ostensible SSE. That is,
\begin{eqnarray}
d\ket{\bar{\psi}_{\bf F}(t)}&=&dt
\Big{(}-\frac{i\Omega}{2}\hat{\sigma}_x+[f(t+dt)-\mu](\rt{\gamma}\hat{\sigma}-\mu/2)\nl
-\smallfrac{1}{2} [\gamma\hat{\sigma}\dg\hat{\sigma}
-\rt{\gamma}\mu\hat{\sigma}+\mu^2/4]
\Big{)}
\ket{\bar{\psi}_{\bf F}(t)}.\hspace{.8cm}
\end{eqnarray}
Thus to determine $\rho_{\bf R}(x,t)$ all we need to do is solve
the above SSE and \erfs{Eq.Linear1a}{Eq.Linear3} for ${\bf R}$
assumed known and $f(t+dt)dt=d{\cal W} +dt \mu$ where $d{\cal W}$
is a Wiener increment. Once solved the quantum state conditioned
on $\bfr$ is given by
\begin{eqnarray}
{\chi}_{\bf R}(t)&=&\frac{{\rm E}_{\bf F}\Big{[}p_{\bf R,
F}(t)\check{\chi}_{\bf F}(t)\Big{]}}{ {\rm E}_{\bf F}\Big{[}p_{\bf
R, F}(t)\check{p}_{\bf F}(t)\Big{]}},
\end{eqnarray} where ${\chi}_i =\{\check{x}_i,\check{y}_i,\check{z}_i\}$ are the Bloch vectors of the quantum state.
The moments of the
classical state are given by
\begin{eqnarray}
\langle x^m_{\bf R} \rangle_{t}&=&\frac{{\rm E}_{\bf
F}\Big{[}{x}^m_{\bf F}(t)p_{\bf R, F}(t)\check{p}_{\bf
F}(t)\Big{]}}{ {\rm E}_{\bf F}\Big{[}p_{\bf R, F}(t)\check{p}_{\bf
F}(t)\Big{]}} . \hspace{.5cm}
\end{eqnarray}

\begin{figure}\begin{center}
\includegraphics*[width=0.45\textwidth]{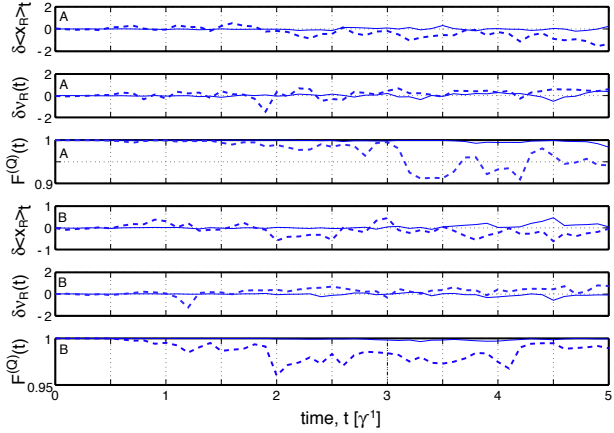}
\caption{\label{figsuper2} This figure shows the quantum and
classical fidelity between the actual solution and our ostensible
solution for ${\rho}_{\bfr}(x,t)$.  Part A corresponds to the
$\lambda=\mu=0$  case while part B represents the $\lambda=0$ and
$\mu=\rt{\gamma}{\rm Tr} [(\hat \sigma+\hat\sigma\dg)\rho_{\bf
F}(t)]$ case. In both case an ensemble size of $n=100$ (dotted)
and $n=10 000$ (solid) was used. The system parameters are the
same as in Fig. \ref{figsuperI}.}\end{center}
\end{figure}

To illustrate this method we considered two choices for the
ostensible distributions. The first is $\lambda=\mu=0$; that is,
all the ostensible distributions are Gaussian distributions of
mean zero and variance $dt$. The second case corresponds to the
situation when $\lambda=0$ and $\mu=\rt{\gamma}{\rm Tr} [(\hat
\sigma+\hat\sigma\dg)\rho_{\bf F}(t)]$; that is, the fictitious
distribution is treated as the real unobservable distribution.
Both cases were simulated to show the robustness of our numerical
technique and to demonstrate that while any ostensible
distributions can be chosen a more realistic choice will result in
a faster convergence. To demonstrate this we numerically solved \erf{supertrajextory}
and used this as our reference solution. Then we compared the mean
and variance of the classical marginal states
and the fidelity for quantum reduced states
(using \erf{felquant} once the
classical space has been removed) for both ostensible cases and with ensemble sizes
of 100 and 10 000.
These results are shown Fig. \ref{figsuper2} where it is observed
that for the larger ensemble size the difference in the classical marginal state
is small and the quantum
fidelity is close to one, indicating that our ostensible method
has reproduced the known result and is converging. Furthermore it
is observed that for the second case for the same ensemble size
this difference is smaller thereby indicating that the second method
convergence is faster.

\section{Discussion and Conclusion}\label{con}

The central topic of this paper was to investigate the conditional
dynamics of partially observed systems (classical and quantum).
Due to the fact that the information obtained is incomplete we have
to assign a mixed state to the system. For a quantum system this
means the state of knowledge given result $r$ is given by the
state matrix $\rho_{r}(t)$ and for a classical system a
probability distribution $P_{r}(x,t)$ has to be used. If we
consider a joint system (for example a classical detector is used
to monitor a quantum system) the conditional state is given by
$\rho_{r}(x,t)$.

Even when we consider continuous-in-time monitoring we can still
have incomplete information because of unobserved processes. For
this case the conditional state trajectories obey either a
stochastic master equation (for a quantum system), a Kushner
Stratonovich equation (for a classical system) or a superoperator
Kushner-Stratonovich equation (for the joint system). That is, to
simulate the conditional state we have to solve a rather
numerically expensive equation. In this paper we showed that by
introducing a fictitious record $\bff$ for the unobserved
processes and ostensible measurement theory we can reduce this
problem to solving pure states (stochastic \sch~ equations for the
quantum system or stochastic differential equations for the
classical system) conditioned on both $\bfr$ and $\bff$. Then by
averaging over all possible $\bff$ we get the require conditional
state. That is the numerical memory requirements are decreased by
a factor of $N$, the number of basis states for the system.
However, this is at the cost of an ensemble average.

In summary, our ostensible method will be useful for investigating
realistic situations where the dimensions of the systems are
large. It is also much easier to implement numerically than the standard
technique, so we expect
it to find immediate applications.

\acknowledgments

We would like to acknowledge the interest shown and help provided
by K. Jacobs and N. Oxtoby. This work was supported by the
Australian Research Council (ARC) and the State of Queensland.

\appendix

\section{Why it is necessary to use the ostensible method}\label{AppendixQuant}

To show that we must use an ostensible distribution rather then
the real distribution, for our numerical technique, we consider
two consecutive measurements. The state of the system (which we take to be quantum for specificity)  after these
two measurement is
\begin{eqnarray}
\rho_{r_2,r_1}&=&\sum_{f_2,f_1}
P(f_2,f_1|r_2,r_1)\rho_{r_2,f_2,r_1,f_1}\nn\\
&=& \frac{\sum_{f_2,f_1}\hat{\cal
J}[\hat{M}_{f_2,r_2}\hat{M}_{f_1,r_1}]\rho(0)}{P(r_1,r_2)}.
\end{eqnarray} This can be rewritten as
\begin{equation}\label{A2}
\rho_{r_2,r_1}
= \sum_{f_2,f_1}\frac{\hat{\cal J}[\hat{M}_{f_2,r_2}]\hat{\cal
J}[\hat{M}_{f_1,r_1}]\rho(0)}{P(r_2,f_2|r_1,f_1)P(r_1,f_1)}\frac{P(r_2,f_2,r_1,f_1)}{P(r_1,r_2)}.
\end{equation}
The first term can be viewed  as the part that determines the
trajectory and the second term as the part which determines the
weighting factor for this trajectory. Considering only the
weighting factor we can rewrite this as $P(f_2,f_1|r_2,r_1)$, which,
unless we have the full numerical solution, is not determinable. To
be more specific we cannot separate this term into $P(f_1)
P(f_2|f_1,r_1)$ and thus we cannot create a trajectory that steps
through time with the correct statistics for $f_k$.

However by introducing an ostensible distribution we can rewrite
\erf{A2} as
\begin{eqnarray}
\rho_{r_2,r_1}&=& \sum_{f_2,f_1}\bar\rho_{f_2,r_2,f_1,r_1}\frac{\Lambda(f_2,r_2,
f_1,r_1)}{P(r_1,r_2)},
\end{eqnarray}
where\begin{equation} \bar\rho_{f_2,r_2,f_1,r_1}= \frac{\hat{\cal
J}[\hat{M}_{f_2,r_2}\hat{M}_{f_1,r_1}]\rho(0)}{\Lambda(f_2,r_2|
f_1,r_1)\Lambda(f_1,r_1)}
\end{equation} and we have complete freedom to choose any $\Lambda(f_2,r_2,
f_1,r_1)$. As such we are not restricted to using the undeterminable 
distribution $P(f_2,r_2,f_1,r_1)$.
This implies that to unravel the conditional state conditioned on some real record ${\bf R}$
in terms of fictitious results the corresponding
trajectories must be unnormalized.

\begin{figure}\begin{center}
\includegraphics*[width=0.45\textwidth]{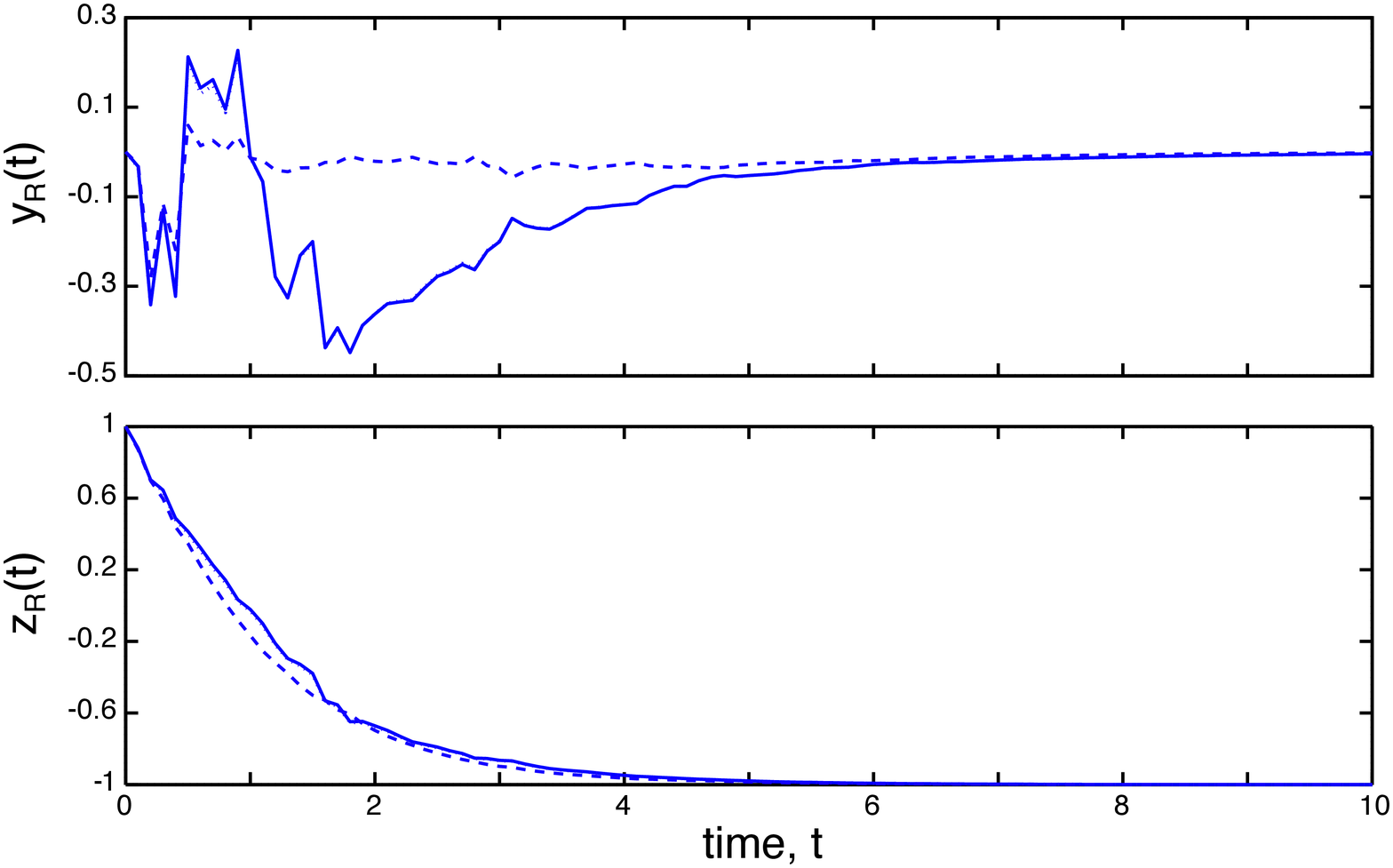}
\caption{\label{AppA} This figure shows a typical trajectory for a radiatively
damped TLA undergoing homodyne-$x$ detection of efficiency $0.4$ (solid line). The dotted line
represents the solution using our method whereas the dashed line represents the BG method.
Both our method and the BG method were implemented using an ensemble size of 10000.}\end{center}
\end{figure}

\begin{figure}\begin{center}
\includegraphics*[width=0.45\textwidth]{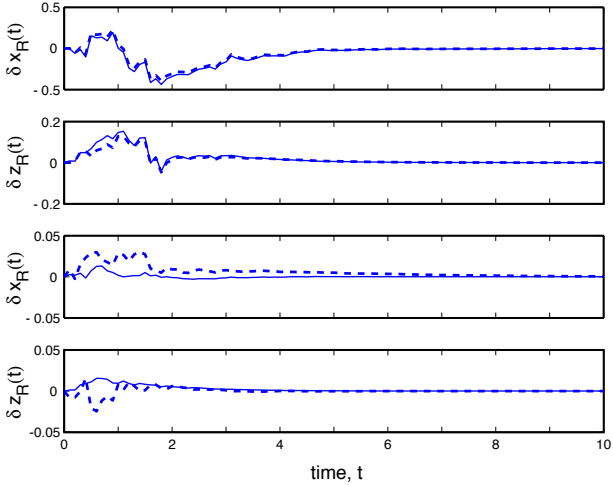}
\caption{\label{AppB} The top two plots show the difference between the exact solution and the BG method for $n=1000$ and 10000.
The bottom two plots show the difference between the exact solution and our  method for $n=1000$ and 10000.}\end{center}
\end{figure}

In this paper we make two choices for $\Lambda(f_2,r_2,
f_1,r_1)$. The first choice is $\Lambda(f_2,r_2,
f_1,r_1) = \Lambda(f_2)\Lambda(f_1)\Lambda(r_2)\Lambda(r_1)$ where
$\Lambda(r_k)$ and $\Lambda(f_k)$ are Gaussian
distributions of variance $dt$ and mean $\lambda$ and $\mu$
respectively. The second choice is  $\Lambda(f_2,r_2,
f_1,r_1)=\Lambda(f_2|f_1,r_1)\Lambda(f_1)\Lambda(r_2)\Lambda(r_1)$ where
$\Lambda(r_k)$ is the same as before but the fictitious results were chosen
based on the true probability we would expect based on the past real and fictitious results up to but not
including the current time. That is
$\mu=\rt{\gamma_2}{\rm Tr[\hat{x}_2\bar\rho_{\bf R , F}(t)]}/{\rm Tr}[\bar\rho_{\bf R, F} (t)]$.

A third possible choice would be  $\Lambda(f_2,r_2,
f_1,r_1)=\Lambda(f_2|f_1,r_1)\Lambda(f_1)\Lambda(r_2|r_1, f_1)\Lambda(r_1)$, where
both the real and fictitious distribution are chosen based on the past results up to but not including
the current time. That is
$\lambda=\rt{\gamma_1}{\rm Tr[\hat{x}_1\bar\rho_{\bf R , F}(t)]}/{\rm Tr}[\bar\rho_{\bf R, F} (t)]$ and
$\mu=\rt{\gamma_2}{\rm Tr[\hat{x}_2\bar\rho_{\bf R , F}(t)]}/{\rm Tr}[\bar\rho_{\bf R, F} (t)]$. This would seem  the
closest choice to $P(f_2,r_2,f_1,r_1)$ but it is important to note that it is not the same. It is still an ostensible distribution, because the true distribution is based upon the entire measurement record, including results in the future. Thus our trajectory equations will still be unnormalized and we cannot replace $(r-\lambda) dt$ with $dW(t)$ as
${\rm Tr}[\hat{x}_1\bar\rho_{\bf R, F}(t)]/{\rm Tr}[\bar\rho_{\bf R, F} (t)]$ is not equal to ${\rm
Tr[\hat{x}_1\rho_{\bf R}(t)]}$ for all possible fictitious records. If we were to make this
substitution we would generate normalized equations,  but  averaging over all possible fictitious record would not give a typical trajectory for the state conditioned on the partial record ${\bf R}$.
This is precisely the mistake Brun and Goan (BG) make in reference \cite{Brun}.

To be more specific let us consider their approach and our approach for the following simple system: A two level atom radiatively damped and monitored using homodyne-$x$ detection with an efficiency  $\eta$. That is this system is
described by the SME
\begin{equation}\label{mainApp}
d\rho_{\bf R}(t)=dt\hat{\cal D}[\hat\sigma]\rho_{\bf R} +\rt{\eta}\hat{\cal H}[\hat\sigma]\rho_{\bf R} dW(t),
\end{equation} where $r(t+dt)dt=dW(t)+dt\rt{\eta}\langle \hat\sigma_x\rangle$. 
Now using BG's theory we would
extend this equation to
\begin{eqnarray} \label{BurnApp}
d\rho_{\bf R,F}(t)&=&dt\hat{\cal D}[\hat\sigma]\rho_{\bf R,F} +\rt{\eta}\hat{\cal H}[\hat\sigma]\rho_{\bf R,F} dW(t)\nl+
\rt{1-\eta}\hat{\cal H}[\hat\sigma]\rho_{\bf R,F} d{\cal W}(t),
\end{eqnarray} which has a pure state solution. 
Now they argue that by ensemble averaging over the fictitious noise
process $d{\cal W}$ \erf{mainApp}
is recovered. However this is incorrect. Although 
\begin{equation}
{\rm E}_{\cal W}[\rt{\eta}\hat{\cal H}[\hat\sigma]\rho_{\bf R,F} dW(t)]
= \rt{\eta} {\rm E}_{\cal W}[\hat{\cal H}[\hat\sigma]\rho_{\bf R,F}] dW(t) ,
\end{equation}
the nonlinearity in the superoperator $\hat{\cal H}$ means that 
\begin{equation}
 {\rm E}_{\cal W}[\hat{\cal H}[\hat\sigma]\rho_{\bf R,F}]  \neq  \hat{\cal H}[\hat\sigma] {\rm E}_{\cal W}[\rho_{\bf R,F}] = \hat{\cal H}[\hat\sigma]\rho_{\bf R}.
\end{equation}

To see this explicitly  we expand \erf{BurnApp} for two measurement (two steps in time)
\begin{eqnarray}
&&\rho_{ r_1,r_2,f_1,f_2}(2dt)=\nl\hspace{1cm}\rho_0+dt\hat{\cal D}[\hat\sigma]\rho_{r_1,f_1} +\rt{\eta}\hat{\cal H}[\hat\sigma]\rho_{r_1,f_1} dW_2(t)+
\nl\hspace{1cm}\rt{1-\eta}\hat{\cal H}[\hat\sigma]\rho_{r_1,f_1} d{\cal W}_2(t)
+dt\hat{\cal D}[\hat\sigma]\rho_{0} + \nl\hspace{1cm} \rt{\eta}\hat{\cal H}[\hat\sigma]\rho_{0} dW_1(t) +
\rt{1-\eta}\hat{\cal H}[\hat\sigma]\rho_{0} d{\cal W}_1(t).
\hspace{0.8cm}
\end{eqnarray} Looking at this equation we see that the problem term is
\begin{eqnarray}
{\rm  E}_{f_1}[\hat{\cal H}[\hat\sigma]\rho_{r_1,f_1}]&=&{\rm E}_{f_1}[\hat{\sigma}\rho_{r_1,f_1}+\rho_{r_1,f_1}\hat\sigma\nl-{\rm Tr}
[\hat\sigma_x\rho_{r_1,f_1}]\rho_{r_1,f_1} ]\nn\\
&\neq&\hat{\sigma}\rho_{r_1}+\rho_{r_1}\hat\sigma-{\rm Tr}
[\hat\sigma_x\rho_{r_1}]\rho_{r_1}
\hspace{0.8cm}
\end{eqnarray} due to the non-linearity.

In our method this problem does not occur because we use ostensible distributions and linear equations. For this system the unnormalized state is
\begin{eqnarray}
d\bar\rho_{\bf R,F}(t)&=&dt\hat{\cal D}[\hat\sigma]\bar\rho_{\bf R,F} +\rt{\eta}\bar{\cal H}_\lambda[\hat\sigma]\bar\rho_{\bf R,F}dt (r-\lambda)+
\nl\rt{1-\eta}\bar{\cal H}_\mu[\hat\sigma]\bar\rho_{\bf R,F} d{\cal W}(t),
\end{eqnarray}
where $\bar{\cal H}_\chi[\hat{A}]\rho=\hat{A}\rho+\rho\hat{A}-\chi\rho$ with $\chi$ being either $\lambda$ or $\mu$. If we expand
this to two measurements we find that the problem term does not occur as
\begin{eqnarray}
{\rm E}_{f_1}[\bar{\cal H}[\hat\sigma]\bar\rho_{r_1,f_1} ]&=&{\rm E}_{f_1}[\hat{\sigma}\bar\rho_{r_1,f_1}+\bar\rho_{r_1,f_1}\hat\sigma-
\lambda\bar\rho_{r_1,f_1} ]\nn\\
&=&\hat{\sigma}\rho_{r_1}+\rho_{r_1}\hat\sigma-\lambda\bar\rho_{r_1}.
\end{eqnarray}  

To show the magnitude of the error in BG's approach we numerically solved the SME 
for $\eta=0.4$ using \erf{mainApp}, BG's method and our method. We first randomly generate 
and store a string of $dW$'s,  and use these to generate the true record ${\bf R}$ via \erf{mainApp}. 
For BG's method we use this string of $dW$'s, and generate an ensemble using randomly generate strings
of $d{\cal W}$s. The resultant solution would, according to BG, correspond to the solution of \erf{mainApp}.
For our method, 
we use the true record ${\bf R}$ and again randomly generate strings of $d{\cal W}$s to obtain an ensemble average. 
The results of these simulations  are shown in Figs. \ref{AppA} and \ref{AppB}. Here we see that BG theory disagrees 
significantly with the exact result. Furthermore this discrepancy cannot be a statistical error as Fig. \ref{AppB} 
shows that when the ensemble average
is increased this difference remains approximately constant. That is, this simulation confirms that BG's method
fails to reproduce a typical solution to \erf{mainApp}. By contrast our method reproduces \erf{mainApp} 
to within statistical error.


\end{document}